\documentclass[pre,aps]{revtex4}

\usepackage{times}
\usepackage{amsmath}
\usepackage{graphicx}
\usepackage{amssymb}
\usepackage{xcolor}

\newcommand{\e}{\mathrm{e}}

\renewcommand{\Re}{\mathop{\mathrm{Re}}}
\renewcommand{\Im}{\mathop{\mathrm{Im}}}

\begin{document}

\title{Solitary waves in a two-dimensional nonlinear Dirac equation:
from discrete to continuum}

\author{J. Cuevas-Maraver}
\affiliation{Grupo de F\'{i}sica No Lineal, Departamento de F\'{i}sica Aplicada I,
Universidad de Sevilla. Escuela Polit\'{e}cnica Superior, C/ Virgen de \'{A}frica, 7, 41011-Sevilla, Spain \\
Instituto de Matem\'{a}ticas de la Universidad de Sevilla (IMUS). Edificio Celestino Mutis. Avda. Reina Mercedes s/n, 41012-Sevilla, Spain}

\author{P. G. Kevrekidis}
\affiliation{Department of Mathematics and Statistics, University of Massachusetts
Amherst, Amherst, MA 01003-4515, USA}

\author{A.B. Aceves}
\affiliation{Department of Mathematics, Southern Methodist University, Dallas, Texas 75275, USA}

\author{Avadh Saxena}
\affiliation{Center for Nonlinear Studies and Theoretical Division, Los Alamos National Laboratory, Los Alamos, New Mexico 87545, USA}

\begin{abstract}
In the present work, we explore a nonlinear Dirac equation motivated as the continuum limit of a binary waveguide array model. We approach the problem both from a near-continuum perspective as well as from a  highly discrete one. Starting from the former, we see that the continuum Dirac solitons can be continued for all values of the discretization (coupling) parameter, down to the uncoupled (so-called anti-continuum) limit where they result in a 9-site configuration. We also consider configurations with 1- or 2-sites at the anti-continuum limit and continue them to large couplings, finding that they also persist. For all the obtained solutions, we examine not only the existence, but also the spectral stability through a linearization analysis and finally consider prototypical examples of the dynamics for a selected number of cases for which the solutions are found to be unstable.
\end{abstract}

\maketitle

\section{Introduction}

Optical waveguide arrays~\cite{dnc,moti} constitute one of the settings that have led to numerous recent experimental and theoretical developments
as regards the analysis of wave phenomena in Hamiltonian lattices. Both in this and in the related context of photorefractive crystals, features such as discrete diffraction~\cite{yaron} and its management~\cite{yaron1}, Talbot revivals~\cite{christo2}, $\mathcal{PT}$-symmetry and its breaking~\cite{kip}, as well as discrete solitons~\cite{yaron,yaron2} and vortices~\cite{neshev,fleischer} were not only theoretically predicted but also experimentally observed. Variants of the theme of optical waveguide arrays have involved multi-component models bearing multiple polarizations~\cite{dncm,rudy}, waveguides featuring quadratic (so-called $\chi^2$) nonlinearities~\cite{stege,rudy2}, and the examination of dark-solitonic states~\cite{shandarov,hadi}.
Another theme where extensive related studies have been conducted is the atomic physics realm of Bose-Einstein condensates (BECs) in optical lattices~\cite{ober,konotop1}.

Recently, research on  binary waveguide arrays has been gaining  momentum~\cite{Tran,akyl,jesuskip,Aceves,yannan}. Part of the reason for this is that under suitable limiting conditions, this system can lead to Dirac-like nonlinear equations that are of increasing prominence and wide relevance in diverse physical contexts. These include, among others, bosonic evolution in honeycomb lattices~\cite{Carr1,Carr2} and a growing class of atomically thin two-dimensional (2D) Dirac materials~\cite{diracmat} such as graphene, silicene, germanene and transition metal dichalcogenides~\cite{tmdc}. They also arise when studying light propagation in honeycomb photorefractive lattices (the so-called photonic graphene)~\cite{peleg,ablowitz3,ablowitz4}.
These Dirac settings have been argued to present fundamental differences, e.g., with respect to their more well known, non-relativistic limits of the nonlinear Schr{\"o}dinger equation, such as the absence of collapse for an extended interval of frequencies in two spatial dimensions; see, e.g., the recent work of~\cite{PRL} and the discussion therein.

Our aim in the present work is to revisit the context of binary waveguide arrays, motivated by the realistic models studied in~\cite{Tran,akyl} for two-dimensional geometries. In an earlier work, we considered the phenomenology of a 1D discrete nonlinear Dirac equation~\cite{JPA}. However, this was done for a discretization of the less straightforwardly applicable (from a physical perspective) Soler model. Here, we turn our attention to the more realistic setting of the waveguide arrays with onsite nonlinearity, aiming to explore existence, spectral stability and dynamics of nonlinear modes. We do this in a complementary way between the (highly) discrete and the continuum limits. On the one hand, we explore the configurations at strong coupling (the continuum soliton) and subsequently reduce the coupling going towards the discrete limit. Here, we eventually find that a configuration bearing nine sites
turns out to be the limiting highly discrete analogue of the continuum solitary wave. On the other hand, we also start our search for model solutions from the highly discrete limit of vanishing coupling (the so called anti-continuum limit of~\cite{macaub}) with one- or two-site configurations and continue them into the strong coupling regime. Utilizing a spectral stability analysis, we identify the regimes of lattice coupling as well as of solution frequency for which the relevant waveforms are dynamically stable. When instability is identified, some prototypical examples of the configuration's unstable evolution are given.

Our exposition will be structured as follows: in Sec. II we provide an overview of the theoretical properties of both the discrete and the continuous
model. In Sec. III, we numerically explore the existence, stability and dynamical features of the models, while in Sec. IV we summarize our findings and present our conclusions as well as some interesting directions for future work.

\section{Theoretical Setup}

Following the setup of \cite{akyl,Tran}, the two-dimensional (continuum) Dirac model of relevance to the binary waveguide problem is of the form:
\begin{eqnarray}\label{eq:dyncont}
    i\partial_t\psi_1 &=& -(i\partial_x+\partial_y)\psi_2+(m-g|\psi_1|^2)\psi_1 ,\nonumber \\
    i\partial_t\psi_2 &=& -(i\partial_x-\partial_y)\psi_1-(m+g|\psi_2|^2)\psi_2 ,
\end{eqnarray}
where $(\psi_1,\psi_2)$ denotes the spinor field (mode amplitude), while $m$ represents the mass associated with the propagation mismatch between the two
different types of waveguides. The cubic nonlinearity stems from the Kerr effect and breaks the Lorentz symmetry (contrary, e.g., to what is the case with the Soler model; cf.~\cite{JPA,PRL}). We consider this model both at the discrete and at the continuum level.

\subsection{Continuous model}

The analysis of the continuum model of Eq.~(\ref{eq:dyncont}) can be performed in radial coordinates, following a procedure similar to the one proposed in \cite{PRL}. This results in the form:

\begin{eqnarray}\label{eq:dynpolar}
    i\partial_t\psi_1 &=& -\e^{-i\theta}\left(i\partial_r+\frac{\partial_\theta}{r}\right)\psi_2+(m-g|\psi_1|^2)\psi_1,
\nonumber \\
    i\partial_t\psi_2 &=& -\e^{i \theta}\left(i\partial_r-\frac{\partial_\theta}{r}\right)\psi_1-(m+g|\psi_2|^2)\psi_2.
\end{eqnarray}

The form of this equation suggests that we look for stationary solutions as $\psi(\vec r,t)=\exp(-i\omega t)\phi(\vec r)$ with

\begin{equation}
    \phi(\vec r)=\left[\begin{array}{c}
    u(r)\e^{i S\theta} \\
    i\,v(r)\e^{i(S+1)\theta}
    \end{array}\right] ,
\label{ansatz}
\end{equation}
where $u(r)$ and $v(r)$ are real-valued. The value $S\in\mathbb{Z}$ can be cast as the vorticity of the first spinor component (the second component in this formulation has vorticity $S+1$). The equation fulfilled by stationary profiles then only depends on $r$, casting the problem into a 1D one:

\begin{eqnarray}\label{eq:statpolar}
     \left(\partial_r+\frac{S+1}{r}\right)v+(m-\omega-gu^2)u &=& 0 , \nonumber \\
    -\left(\partial_r-\frac{S}{r}\right)u-(m+\omega+gv^2)v &=& 0 ,
\end{eqnarray}
with $r>0$. As we discuss in the next section, stationary solutions are sought by numerical means.

In order to capture the linear stability of the stationary solutions, we introduce the following ansatz into (\ref{eq:dynpolar}):

\begin{equation}
\psi(\vec r,t)
=\left[
    \begin{array}{c}
    \left\{u(r)+\delta\left[a_1(r)\e^{i q\theta}\e^{\lambda t}+b^*_1(r)\e^{-i q\theta}\e^{\lambda^* t}\right]\right\}\e^{i S\theta} \\[2ex]
\!  i\left\{v(r)+\delta\left[a_2(r)\e^{i q\theta}\e^{\lambda t}+b_2^*(r)\e^{-i q\theta}\e^{\lambda^* t}\right]\right\}\e^{i(S+1)\theta}
    \end{array}
\!\right]
\e^{-i\omega t},
\end{equation}
and subsequently solve the ensuing [to O$(\delta)$] eigenvalue problem: $\lambda(a_1,a_2,b_1,b_2)^T=\mathcal{M}_q(a_1,a_2,b_1,b_2)^T$ with $\mathcal{M}_q$ being

\begin{equation}\label{eq:stabcont}
    \mathcal{M}_q=i\left(\begin{array}{cc} L_1 & L_2 \\ \\ -L_2^* & -L_1^* \end{array}\right)
    -\frac{iq}{r}\left(\begin{array}{cc} \sigma_1 & 0 \\ \\ 0 & \sigma_1 \end{array}\right) ,
\end{equation}
and
\begin{equation}
    L_1=
    \left(\begin{array}{cc}
    \omega-m+2gu^2 & -\left(\partial_r+\frac{S+1}{r}\right) \\ \\
    \left(\partial_r-\frac{S}{r}\right) & \omega+m+2gv^2
    \end{array}\right)\ ,
\quad
    L_2=
    \left(\begin{array}{cc}
    gu^2 & 0 \\ \\
    0 & gv^2
    \end{array}\right)\ ,
\quad
    \sigma_1=
    \left(\begin{array}{cc}
    0 & I \\ \\
    I & 0
    \end{array}\right)\ .
\end{equation}

The key observation which facilitates a computation of the spectrum is that the explicit form of Eq.~(\ref{eq:stabcont}) for $\mathcal{M}_q$ contains $r$ and $\partial_r$, but not $\theta$. This allows us to compute the full 2D stability spectrum $\mathcal{M}$ as the union of spectra of the one-dimensional spectral problems:

\begin{equation}
    \sigma\left(\mathcal{M}\right)=\bigcup_{q\in\mathbb{Z}}\sigma\left(\mathcal{M}_q\right).
\end{equation}

In what follows, for concreteness we will set $m=g=1$ (as this choice can be made by renormalizing the time and the wave function). Now, the only free parameter that will be considered in the continuum limit is the frequency $\omega$.

\subsection{Discrete model}

The discrete version of Eq.~(\ref{eq:dyncont}) will be based on a discretization similar to that considered, e.g., in~\cite{akyl} (cf. the discussion around Eq. (2) therein). From a numerical approximation perspective, this is tantamount to a centered difference discretization of the first derivative in Eq.~(\ref{eq:dyncont}) and leads to:

\begin{eqnarray}\label{eq:dyn}
    i\partial_t V_{n,m} &=& -C[i\nabla_x U_{n,m}+\nabla_y U_{n,m}]+(m-g|V_{n,m}|^2)V_{n,m} ,\nonumber \\
    i\partial_t U_{n,m} &=& -C[i\nabla_x V_{n,m}-\nabla_y V_{n,m}]-(m+g|U_{n,m}|^2)U_{n,m} ,
\end{eqnarray}
with $U_{n,m}$ and $V_{n,m}$ ($-N/2+1\leq(n,m)\leq N/2$) being the components of the spinor (amplitude modes) $\Psi_{n,m}\equiv(U_{n,m},V_{n,m})$, and $\nabla_x\Psi_{n,m}\equiv(\Psi_{n+1,m}-\Psi_{n-1,m})$, $\nabla_y\Psi_{n,m}\equiv(\Psi_{n,m+1}-\Psi_{n,m-1})$ being the $x$ and $y$ components of the discrete gradient. The connection to the corresponding continuum limit can be assigned by selecting $C=1/(2h)$ with $h$ being the lattice spacing (discretization parameter).

The dynamical system of Eq.~(\ref{eq:dyn}) presents a number of conserved quantities, such as the charge (squared $\ell^2$ norm):
\begin{equation}\label{eq:charge}
    Q=\sum_n\sum_m \rho_{n,m},\qquad \rho_{n,m}=|U_{n,m}|^2+|V_{n,m}|^2,
\end{equation}
with $\rho_{n,m}$ being the charge density, and the Hamiltonian:
\begin{equation}\label{eq:ham}
  H=-\frac{1}{2}\sum_n \left[C V_{n,m}^*(i\nabla_x U_{n,m}+\nabla_y U_{n,m})+
    C U_{n,m}^*(i\nabla_x V_{n,m}-\nabla_y V_{n,m}) -\frac{g}{k+1}(|U_{n,m}|^4+|V_{n,m}|^4)+m(|U_{n,m}|^2-|V_{n,m}|^2)\right].
\end{equation}

Equations (\ref{eq:dyn}) can be derived from the Hamiltonian (\ref{eq:ham}) by means of Hamilton's equations:

\begin{equation}
    \mathrm{i}\dot U_{n,m}=\frac{\delta H}{\delta U_{n,m}^*},\qquad \mathrm{i}\dot V_{n,m}=\frac{\delta H}{\delta V_{n,m}^*} .
\end{equation}

Our main focus hereafter will be on stationary solutions and their stability
as well as their dynamics. Such solutions can be found by using $U_{n,m}(t)=\exp(-i \omega t)u_{n,m}$, $V_{n,m}(t)=\exp(-i \omega t)v_{n,m}$,
when they possess frequency $\omega$ and satisfy the coupled algebraic equations:

\begin{eqnarray}\label{eq:stat}
    (\omega-m+g|v_{n,m}|^2) v_{n,m} +C[i\nabla_x u_{n,m}+\nabla_y u_{n,m}] &=& 0,\nonumber \\
    (\omega+m+g|u_{n,m}|^2) u_{n,m} +C[i\nabla_x v_{n,m}-\nabla_y v_{n,m}] &=& 0.
\end{eqnarray}

Once stationary solutions of the algebraic system of Eqs.~(\ref{eq:stat}) are calculated (by, e.g., fixed point methods as discussed below), their linear stability is considered by means of a linearized stability analysis. More specifically, considering small perturbations [of order ${\rm O}(\delta)$, with $0< \delta \ll 1$] of the stationary solutions, we substitute the ansatz

\begin{equation}
    U_{n,m}(t)=e^{-i \omega t} \left[u_{n,m} + \delta (a_{n,m} e^{\lambda t} + c_{n,m}^{*} e^{\lambda^{*} t}) \right],\ \ \
    V_{n,m}(t)=e^{-i \omega t} \left[v_{n,m} + \delta (b_{n,m} e^{\lambda t} + d_{n,m}^{*} e^{\lambda^{*} t}) \right]
\end{equation}
into Eqs.~(\ref{eq:dyn}), and then solve the ensuing [to O$(\delta)$] eigenvalue problem: $\lambda(a_{n,m},b_{n,m},c_{n,m},d_{n,m})^T=\mathcal{M}(a_{n,m},b_{n,m},c_{n,m},d_{n,m})^T$ with $\mathcal{M}$ being
\begin{equation}\label{eq:stab}
    \mathcal{M}=i\left(\begin{array}{cc} L_1 & L_2 \\ \\ -L_2^* & -L_1^* \end{array}\right)
\end{equation}
and
\begin{equation}
    L_1=
    \left(\begin{array}{cc}
    \omega-m+2g|u|^2 & C(i\nabla_x+\nabla_y) \\ \\
    C(i\nabla_x-\nabla_y) & \omega+m+2g|v|^2
    \end{array}\right)\ ,
\quad
    L_2=
    \left(\begin{array}{cc}
    u^2 & 0 \\ \\
    0 & v^2
    \end{array}\right)\ .
\end{equation}

The potential existence of an eigenvalue with non-vanishing real part suggests the existence of a dynamical instability. If all the eigenvalues are imaginary, then the solution is spectrally (neutrally) stable.

As in the continuum limit, we will set $m=g=1$ and vary $\omega$ as well as $C$ as our relevant parameters in order to characterize the behavior of the solution and the variation of its stability properties.

\section{Numerical results}

\subsection{Continuous model}

\begin{figure}
\begin{tabular}{cc}
\includegraphics[width=6cm]{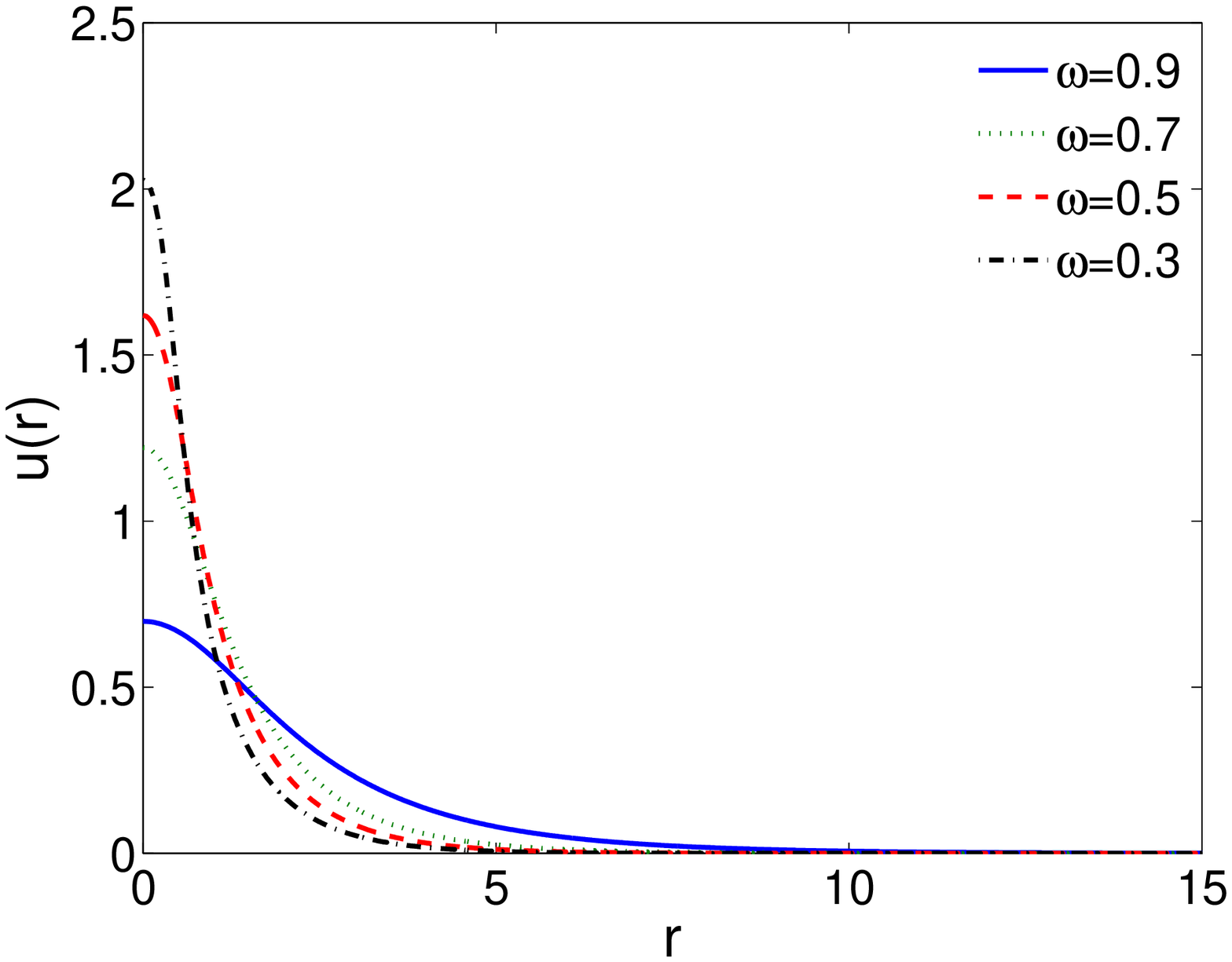} &
\includegraphics[width=6cm]{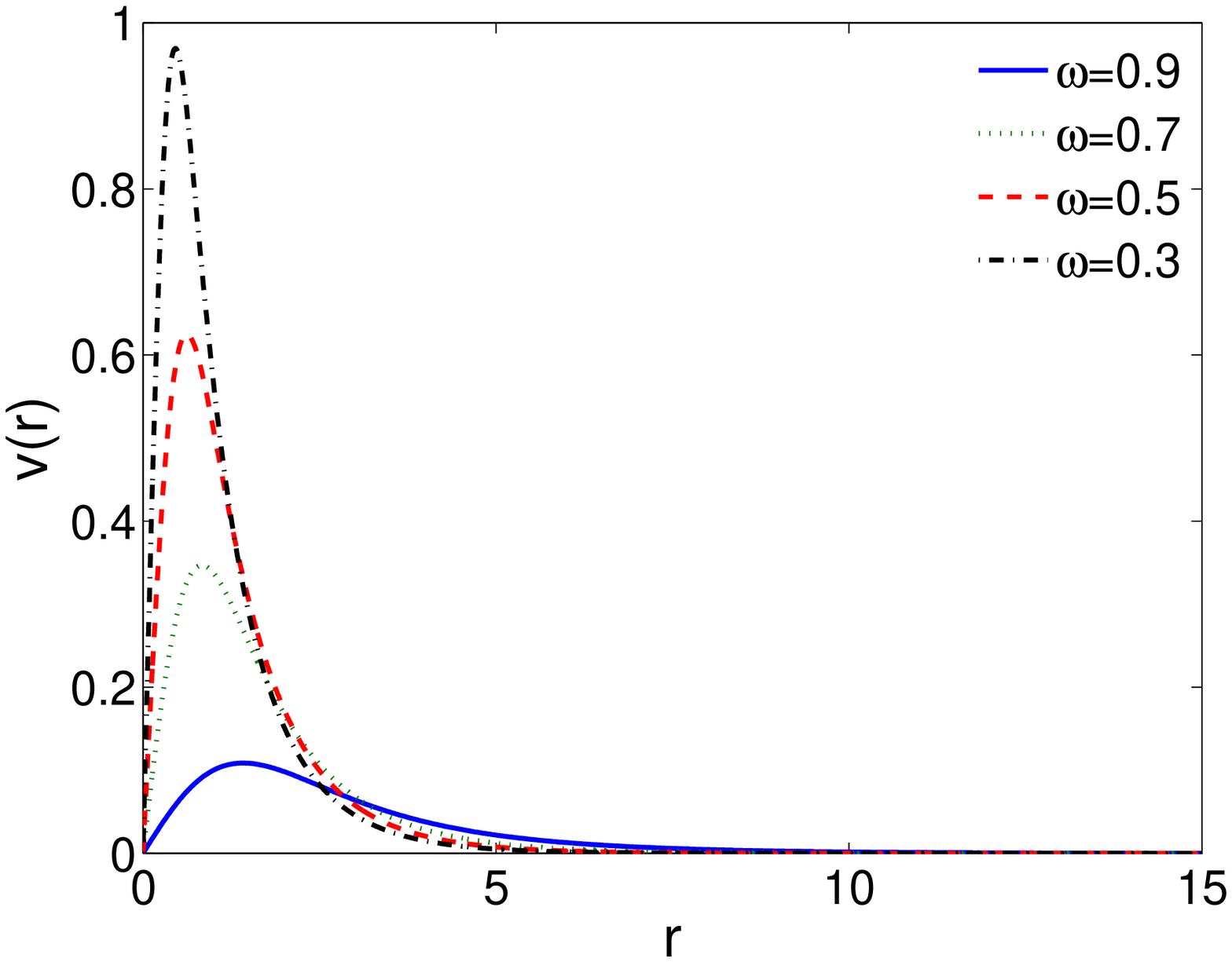} \\
\includegraphics[width=6cm]{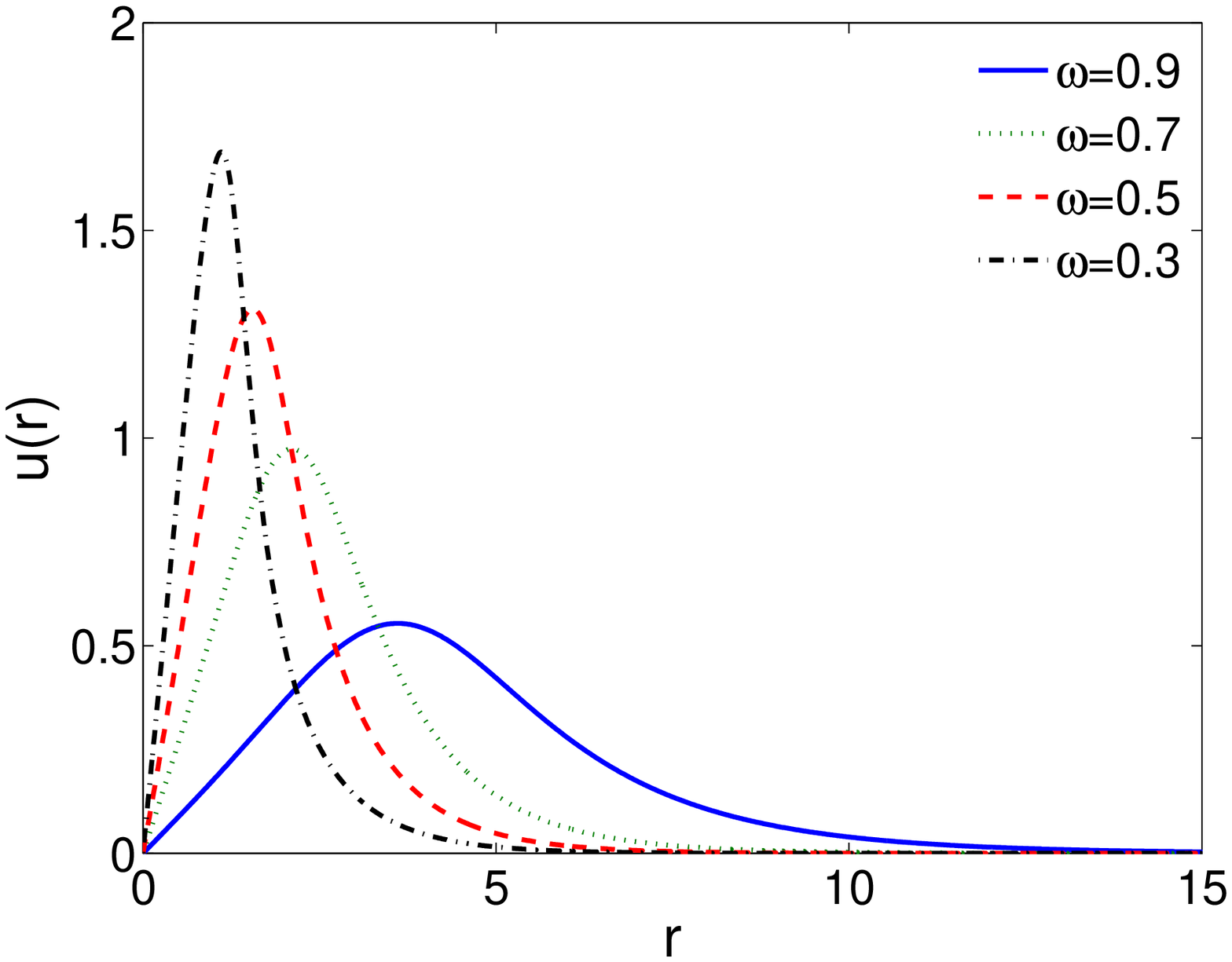} &
\includegraphics[width=6cm]{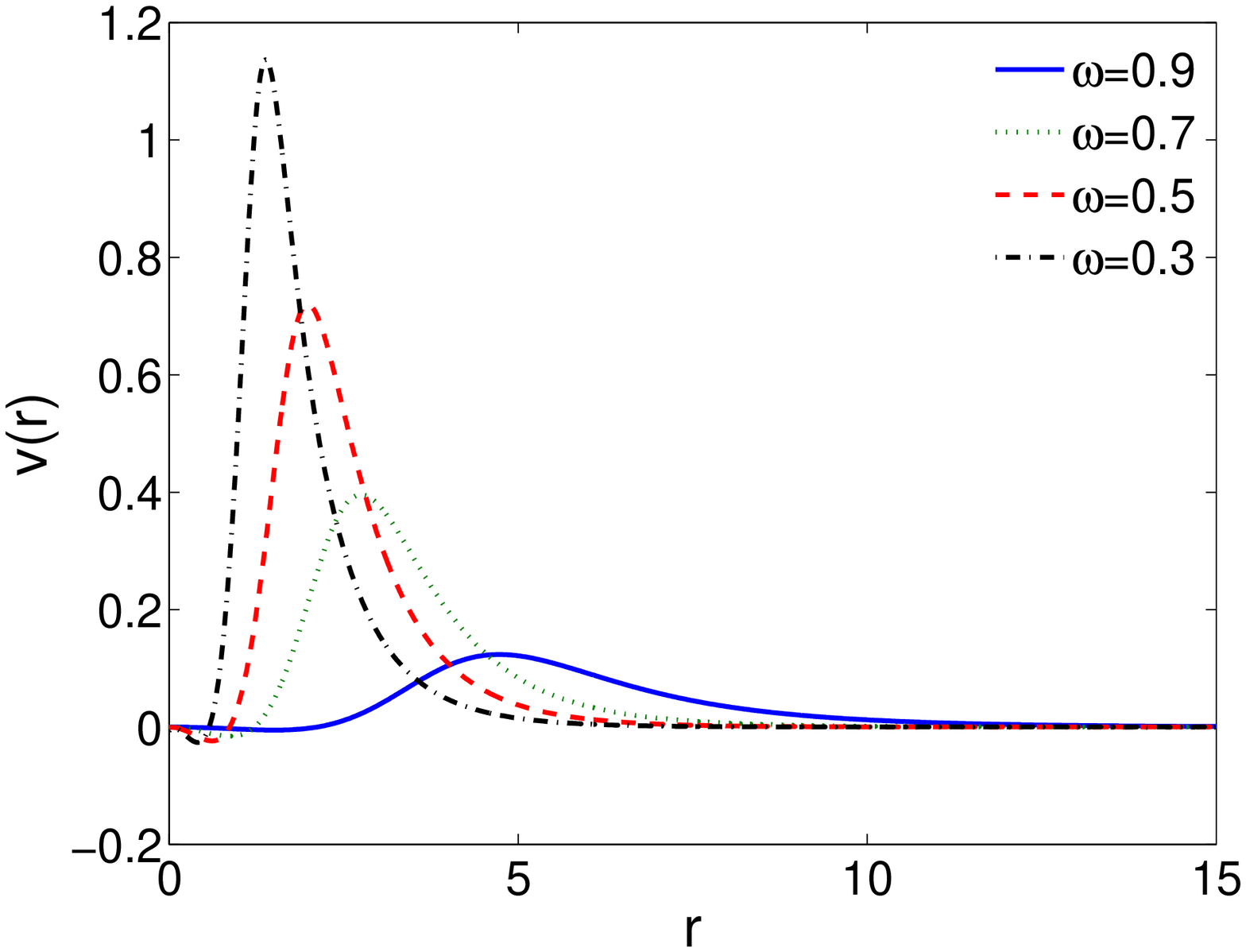} \\
\end{tabular}
\caption{Radial profiles of the spinor components for (upper panels) $S=0$ solitary waves and (lower panels) $S=1$ vortices for different values of $\omega$.}
\label{fig:profilevortex}
\end{figure}

\begin{figure}
\begin{center}
\begin{tabular}{cc}
\includegraphics[width=6cm]{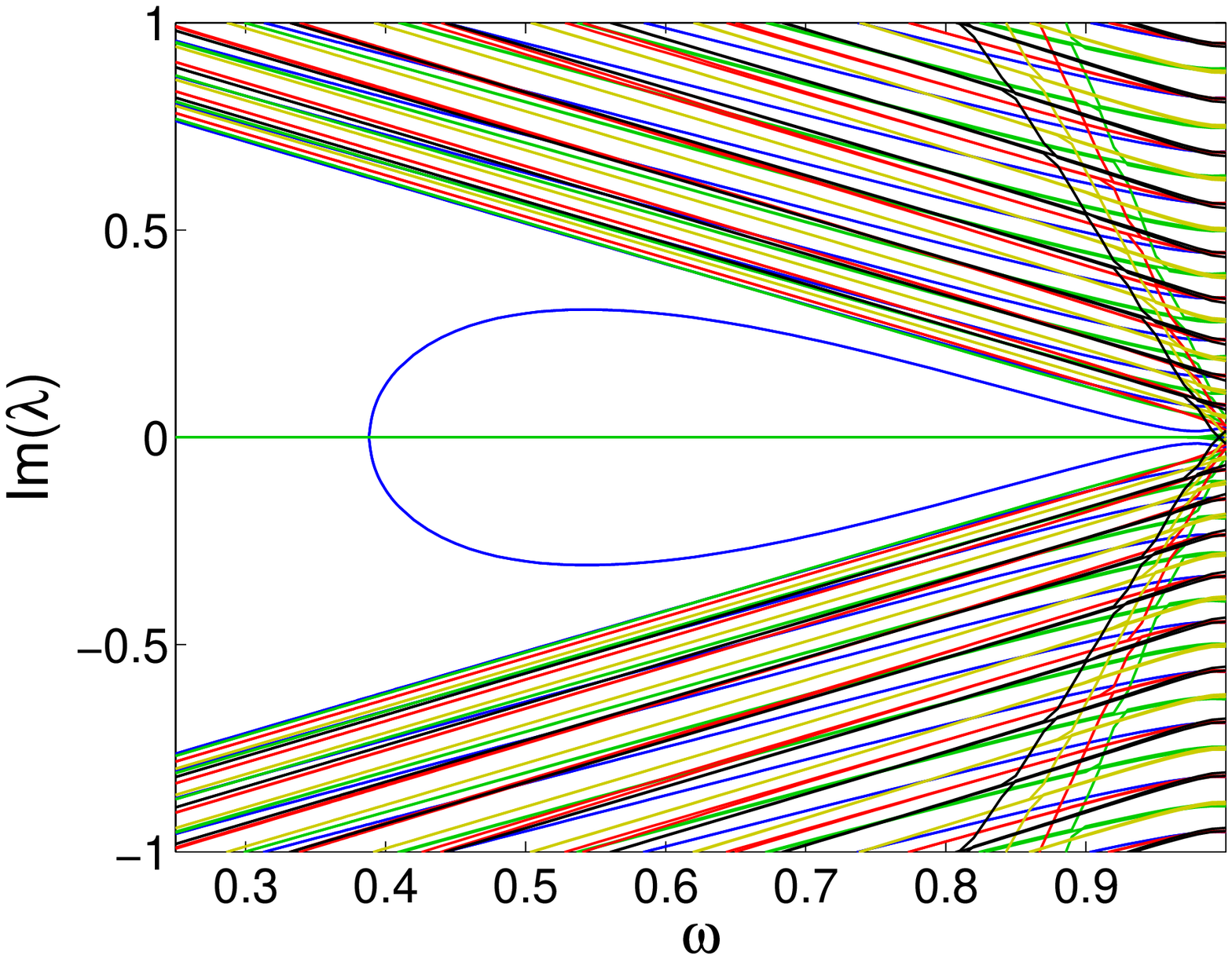} &
\includegraphics[width=6cm]{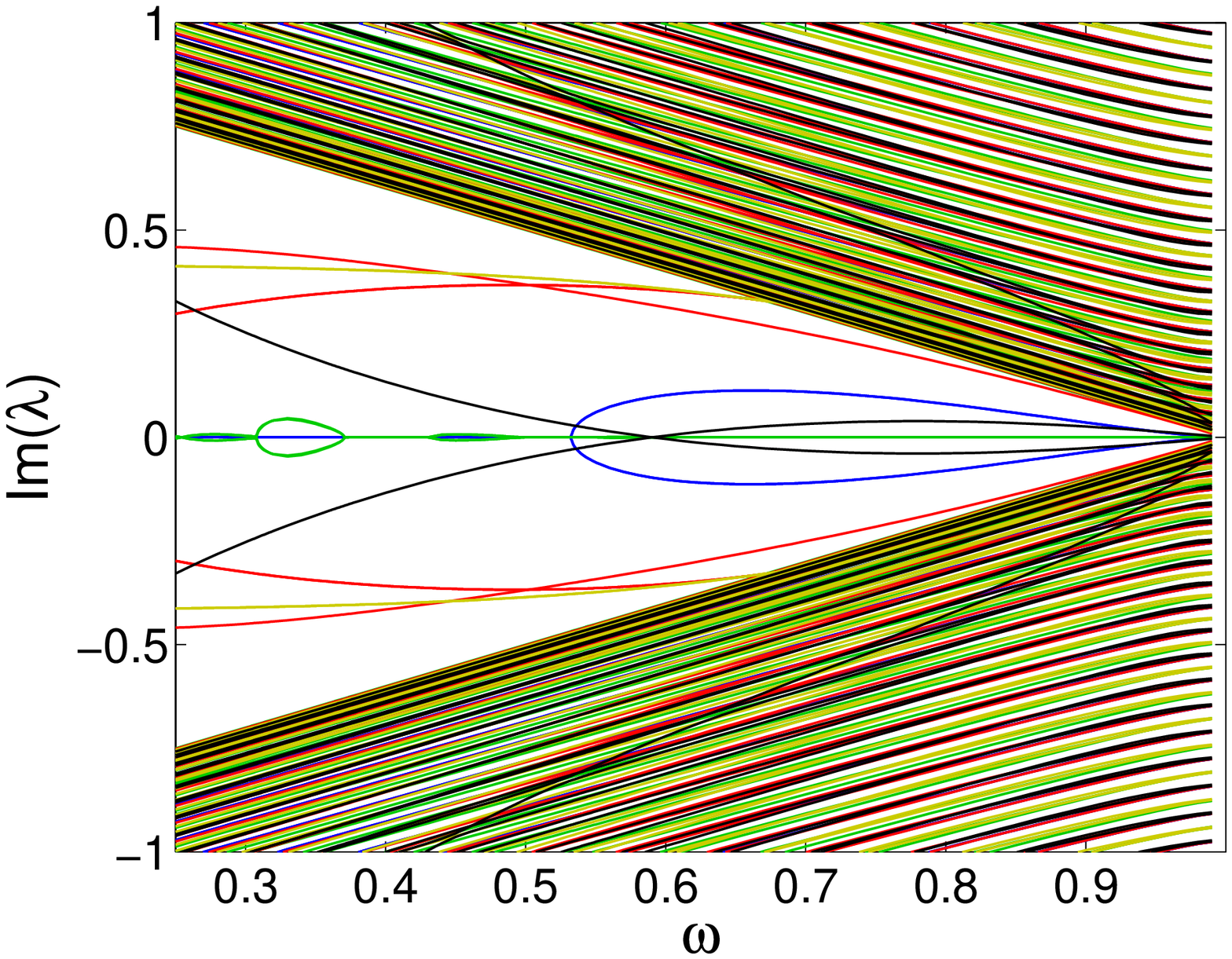} \\
\includegraphics[width=6cm]{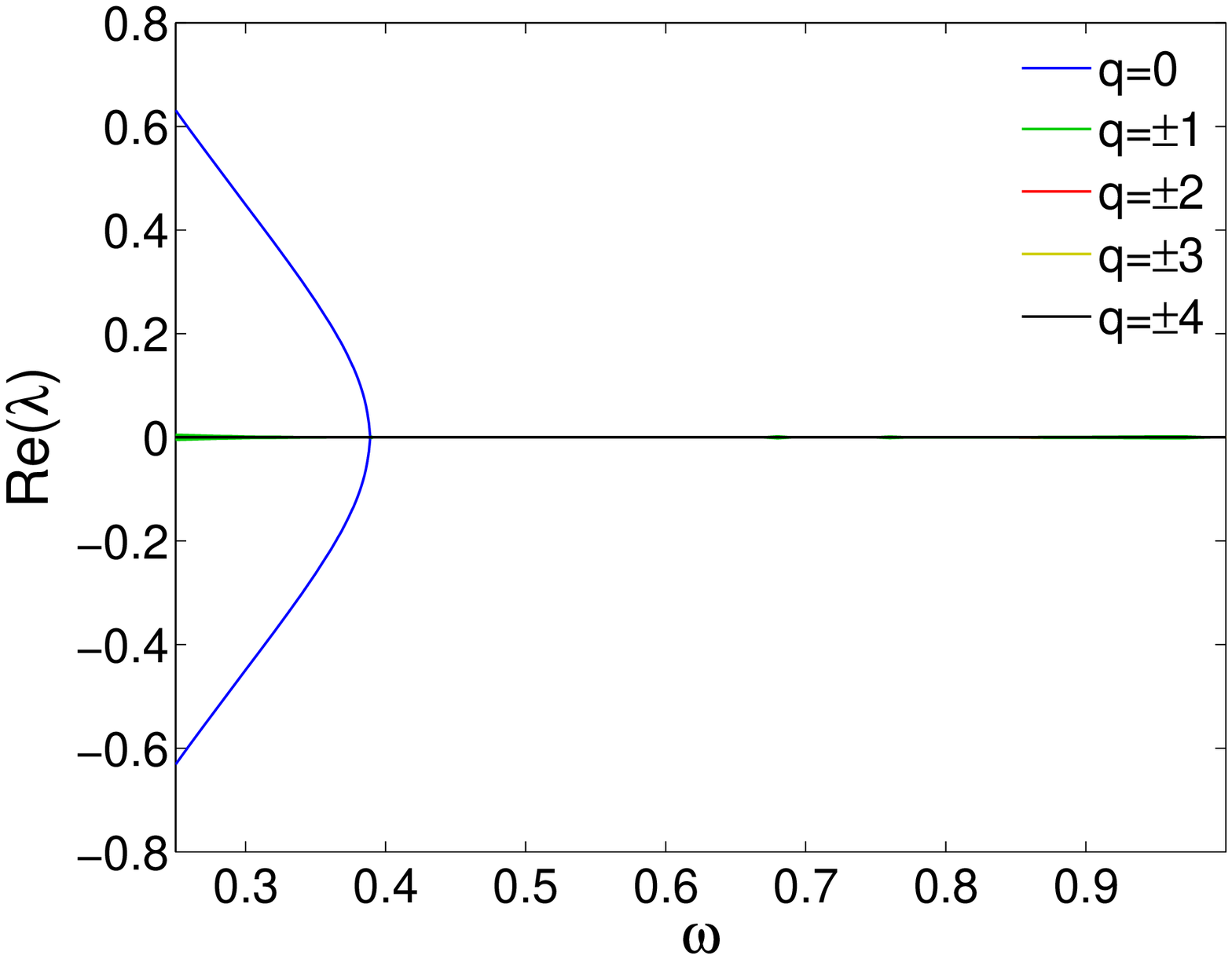} &
\includegraphics[width=6cm]{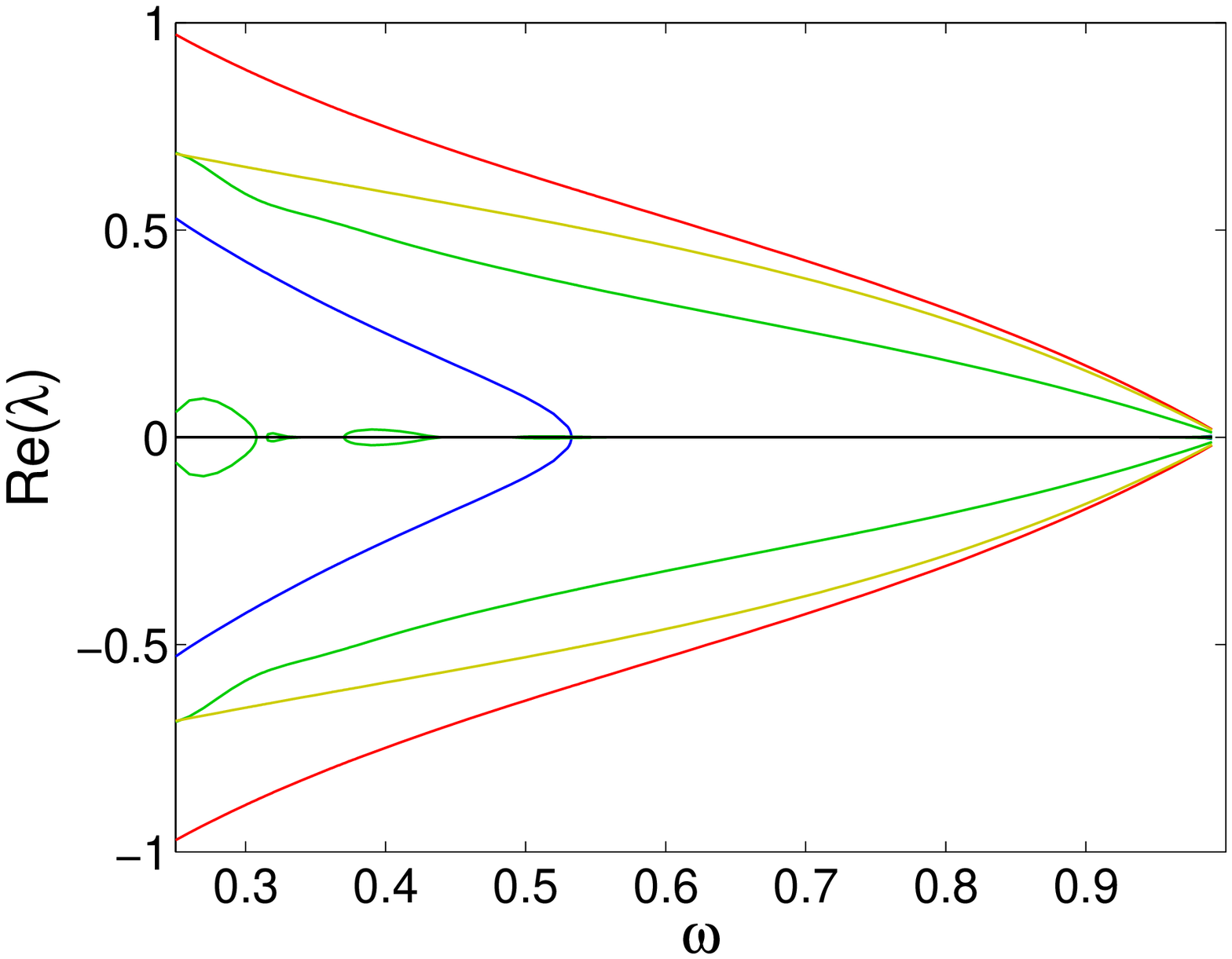} \\
\includegraphics[width=6cm]{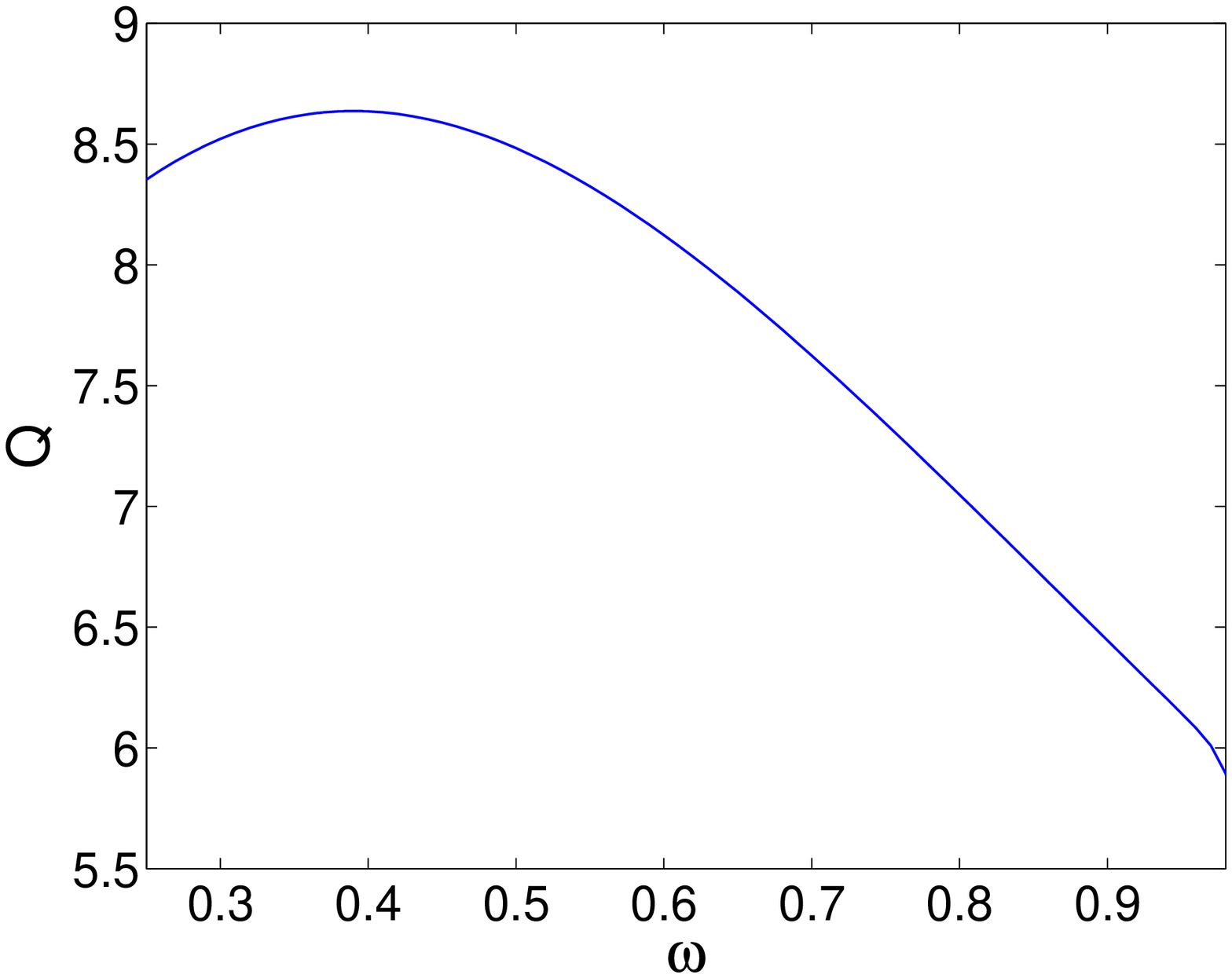} &
\includegraphics[width=6cm]{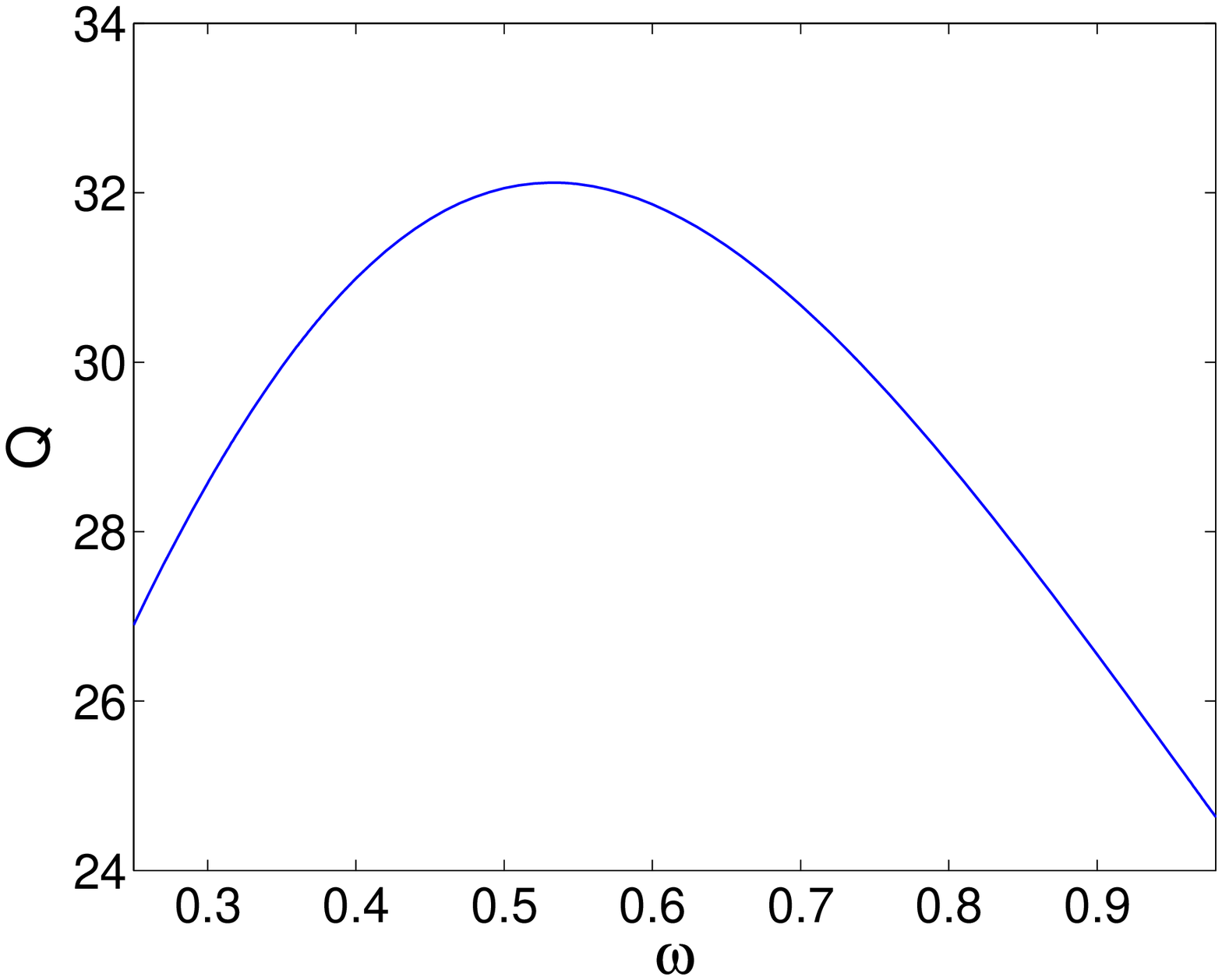} \\
\end{tabular}
\end{center}
\caption{Dependence of the (top) imaginary and (middle) real part of the eigenvalues with respect to $\omega$. Left (respectively, right) panels correspond to $S=0$ solitary waves ($S=1$ vortices). The panels at the left illustrate that the $S=0$ solution is only unstable for $\omega < 0.388$. The panels at the right indicate that the $S=1$ solution is unstable for all values of the frequency $\omega$. The color that corresponds to each $q$ is indicated in the middle left panel. Bottom panels show the dependence of the charge with respect to the frequency. According to the Vakhitov--Kolokolov criterion, the maximum of the curve indicates the occurrence of an exponential instability caused by radially symmetric perturbations ($q=0$).}
\label{fig:spectrumvortex}
\end{figure}

We start our exposition by showing the numerical results regarding
fundamental solutions ($S=0$ solitary waves, for which we nevertheless note that their second
component $\psi_2$ bears a vortex of charge $1$) and $S=1$
vortices in the continuous setting. Numerical analysis has been performed in a similar fashion as in Ref.~\cite{PRL}, using spectral methods for dealing with spatial derivatives. Figure \ref{fig:profilevortex} shows several examples of the profiles for $S=0$ and $S=1$ stationary solutions. To assess stability, Fig.~\ref{fig:spectrumvortex} shows the dependence on the frequency of the real and imaginary parts of the eigenvalues for $S=0$ and $S=1$ solitary waves. We observe that, similar to the Soler model \cite{PRL}, the continuum $S=0$ solitons are unstable below a critical frequency ($\omega=0.388$) for this model. However, it is interesting to observe that contrary to the Soler model, only $q=0$ instabilities are present for the $S=0$ case; these instabilities are of exponential nature and, consequently, can be predicted by the Vakhitov--Kolokolov criterion, as the curve representing charge versus frequency presents a maximum at the bifurcation point {(see bottom panel of Figs. \ref{fig:spectrumvortex})}. In addition, solutions get more localized with decreasing frequency and tend to be localized at $r=0$ as $\omega\rightarrow0$.

\begin{figure}
\begin{center}
\begin{tabular}{cc}
\includegraphics[width=6cm]{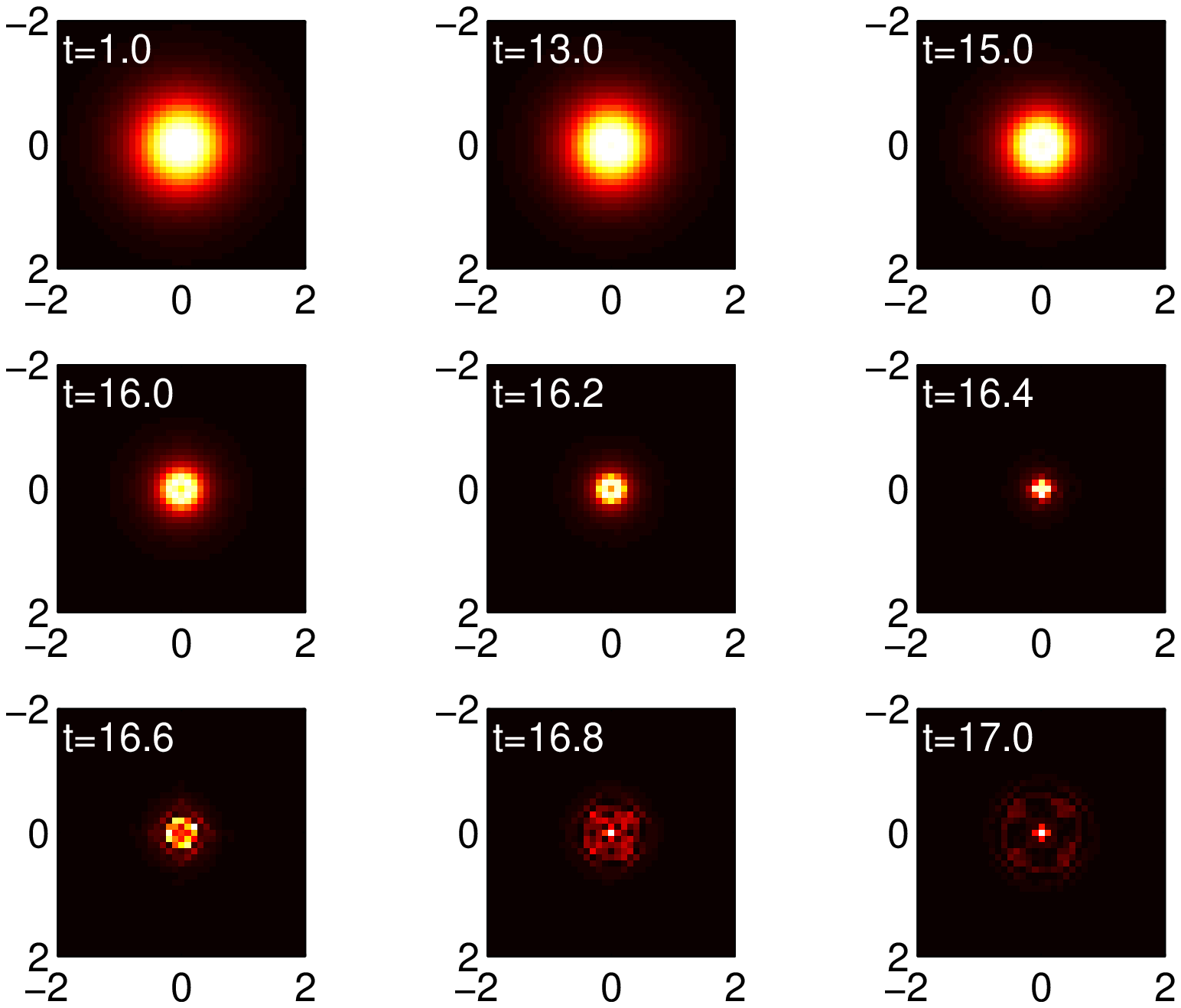} &
\includegraphics[width=6cm]{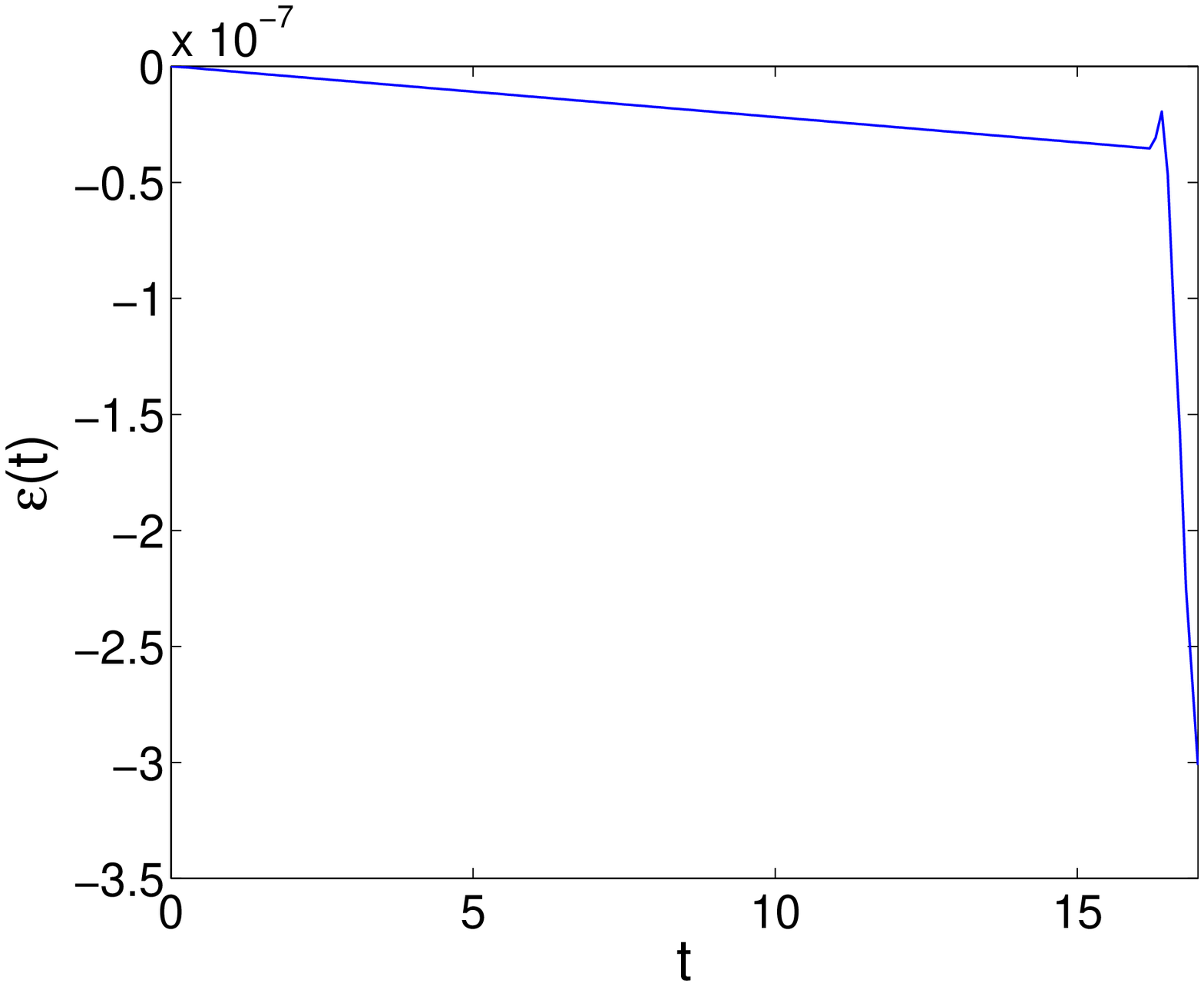} \\
\end{tabular}
\end{center}
\caption{Evolution of the unstable $S=0$ soliton with $\omega=0.3$. Dynamics towards the collapse with the conservation of charge is observed. The left
panels illustrate the evolution of the total density at different times in a two-dimensional contour plot. The right panel shows the evolution of the power based diagnostic of Eq.~(\ref{nnorm}) which is well conserved during the simulation.}
\label{fig:collapse}
\end{figure}

We can see from the stability analysis of Fig.~\ref{fig:spectrumvortex}
that for the fundamental solution, there are wide intervals of
stability/instability for $S=0$, while the $S=1$ waveform
is unstable for all values of $\omega$ (and coupling $C$); cf. with
the similar results of~\cite{PRL}. This
prompts us to turn to the dynamical evolution of the instability.
The study of the dynamics of the unstable solutions for $S=0$ shows that they
feature collapse, as is depicted in Fig.~\ref{fig:collapse}.
As a numerical diagnostic for the accuracy of the simulation,
we have monitored the relative norm error $\varepsilon$ defined as
\begin{eqnarray}
  \varepsilon(t)=\frac{|Q(t)-Q(0)|}{Q(0)} ,
  \label{nnorm}
\end{eqnarray}
with $Q(t)$ being the soliton's charge. We have observed that the norm is preserved within a factor $\sim10^{-7}$, despite an obvious departure of the norm from its initial value as collapse occurs.

In \cite{movie1} one can find a movie with the soliton evolution. In every movie, the top panels correspond to the density of each spinor component and bottom ones to their phase.

\subsection{Discrete model}

We now turn to the existence, stability and dynamics of solitons in the discrete 2D nonlinear Dirac equation (NLDE) in the form of Eq.~(\ref{eq:dyn}).
The solutions are obtained by making use of fixed point methods based on the anti-continuous (AC) limit~\cite{macaub}, that allows to solve Eq.~(\ref{eq:stat}). The principal challenge in this case is to identify a suitable solution in the AC limit given that there are many solutions at $C=0$ that can be extended to the continuous $C\rightarrow\infty$ limit. In fact, all the solutions we have analyzed can be traced upon increasing the coupling strength $C$ towards the continuum limit.

\begin{figure}
\begin{tabular}{ccc}
\includegraphics[width=6cm]{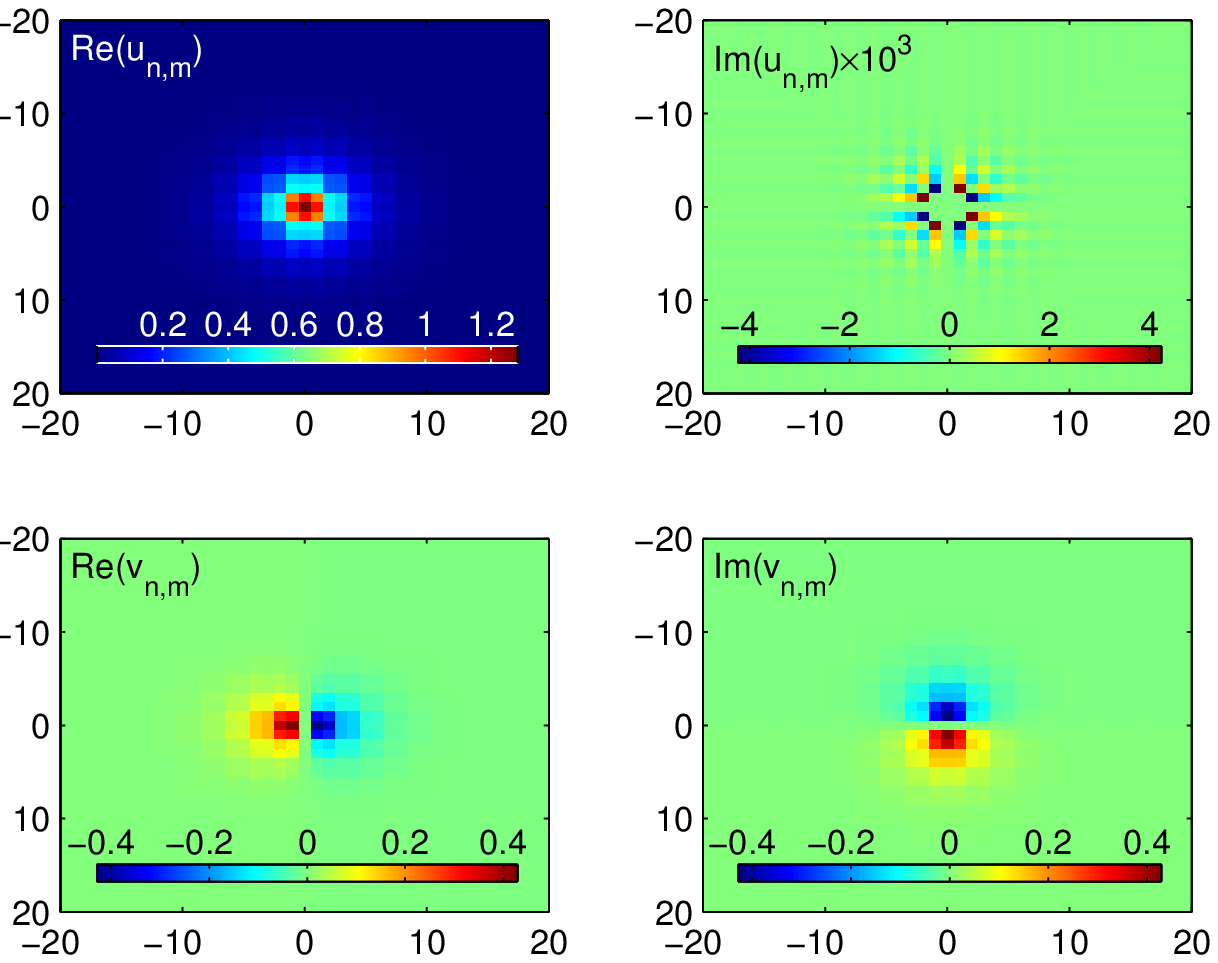} &
\includegraphics[width=6cm]{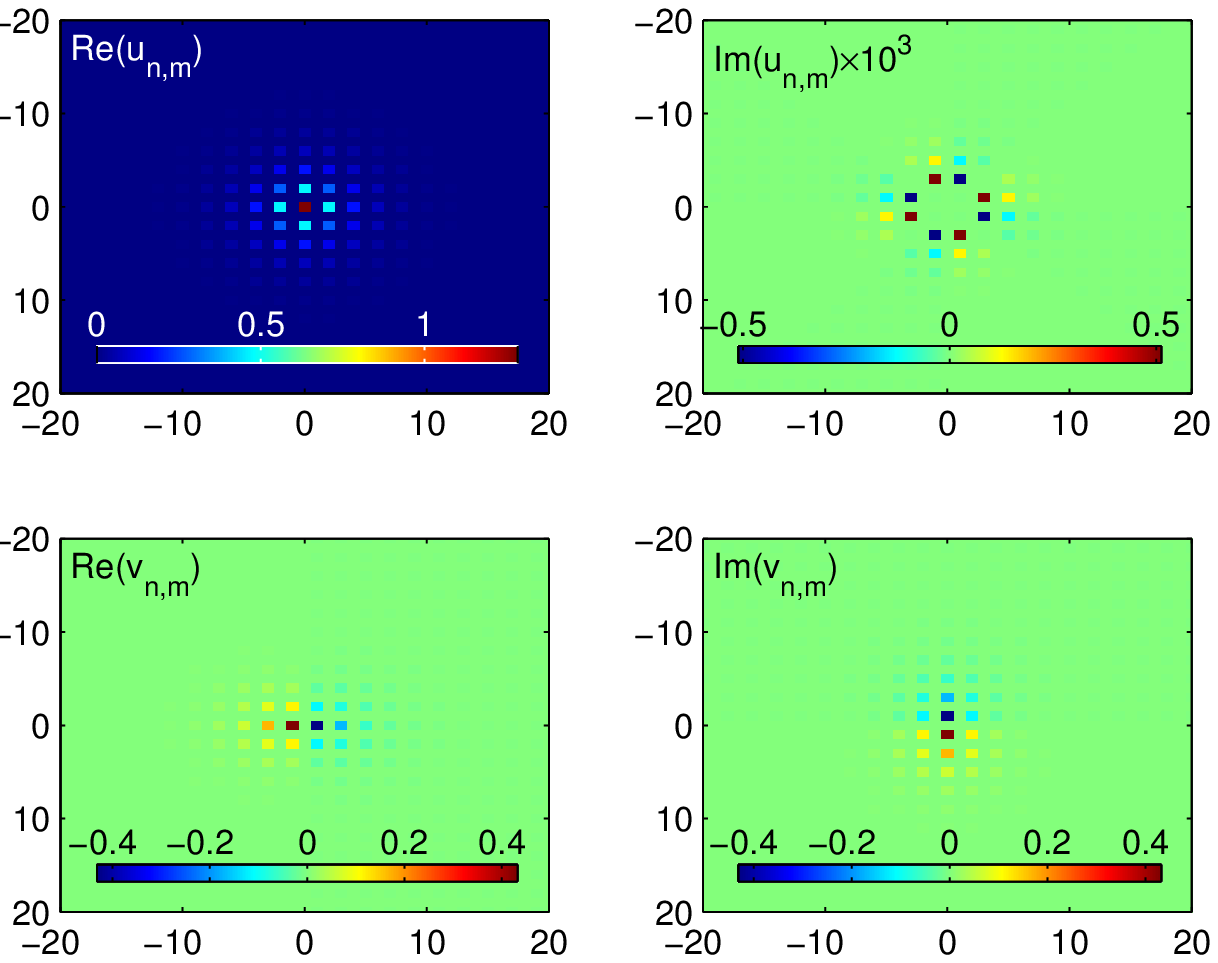} &
\includegraphics[width=6cm]{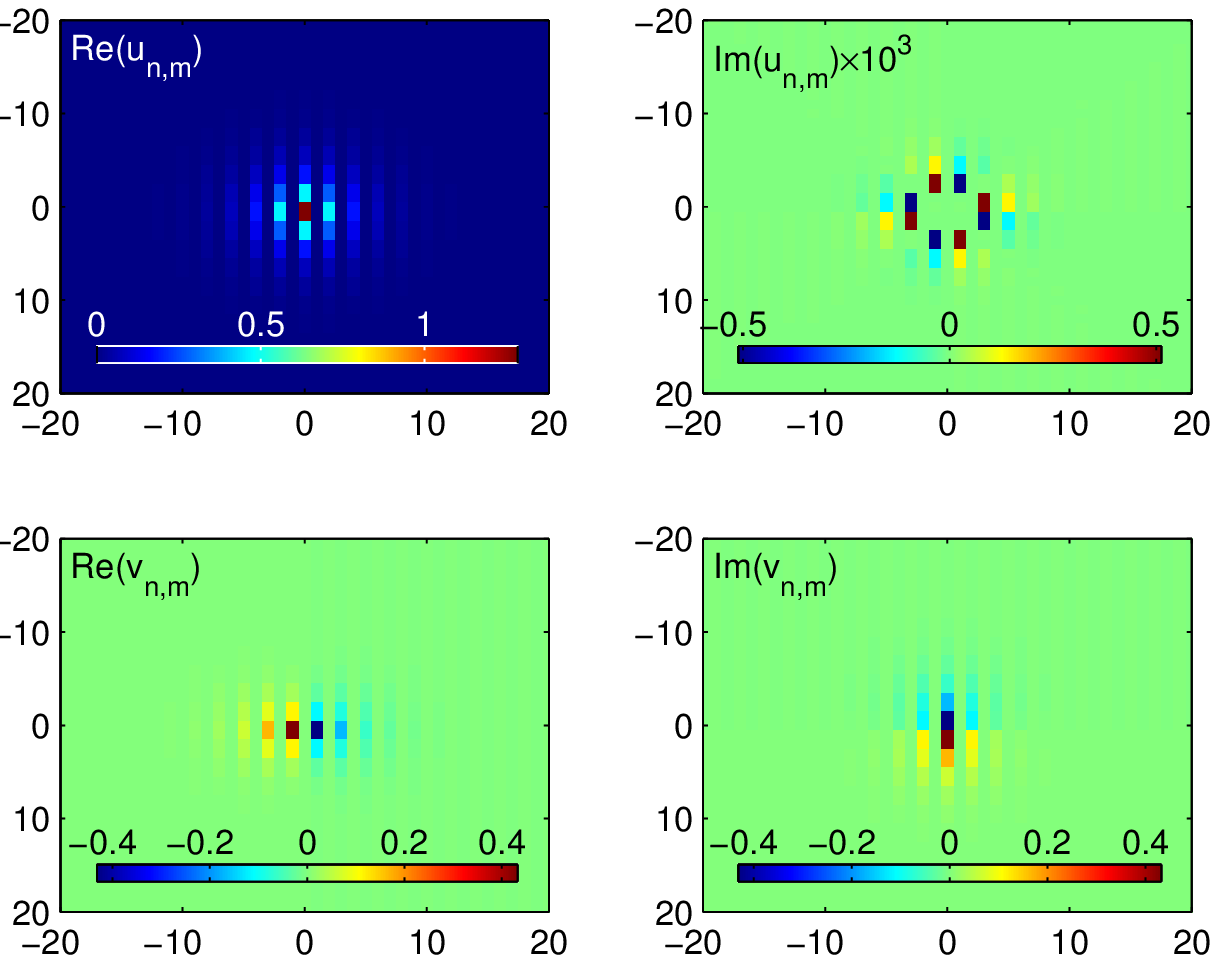} \\
\end{tabular}
\caption{Profiles of the 9-site (left set of $2 \times 2$ panels), 1-site
  (middle set of panels) and 2-site (right set of panels)
  soliton with $\omega=0.7$ and $C=1$.}
\label{fig:profs}
\end{figure}

\begin{figure}
\begin{tabular}{ccc}
\includegraphics[width=6cm]{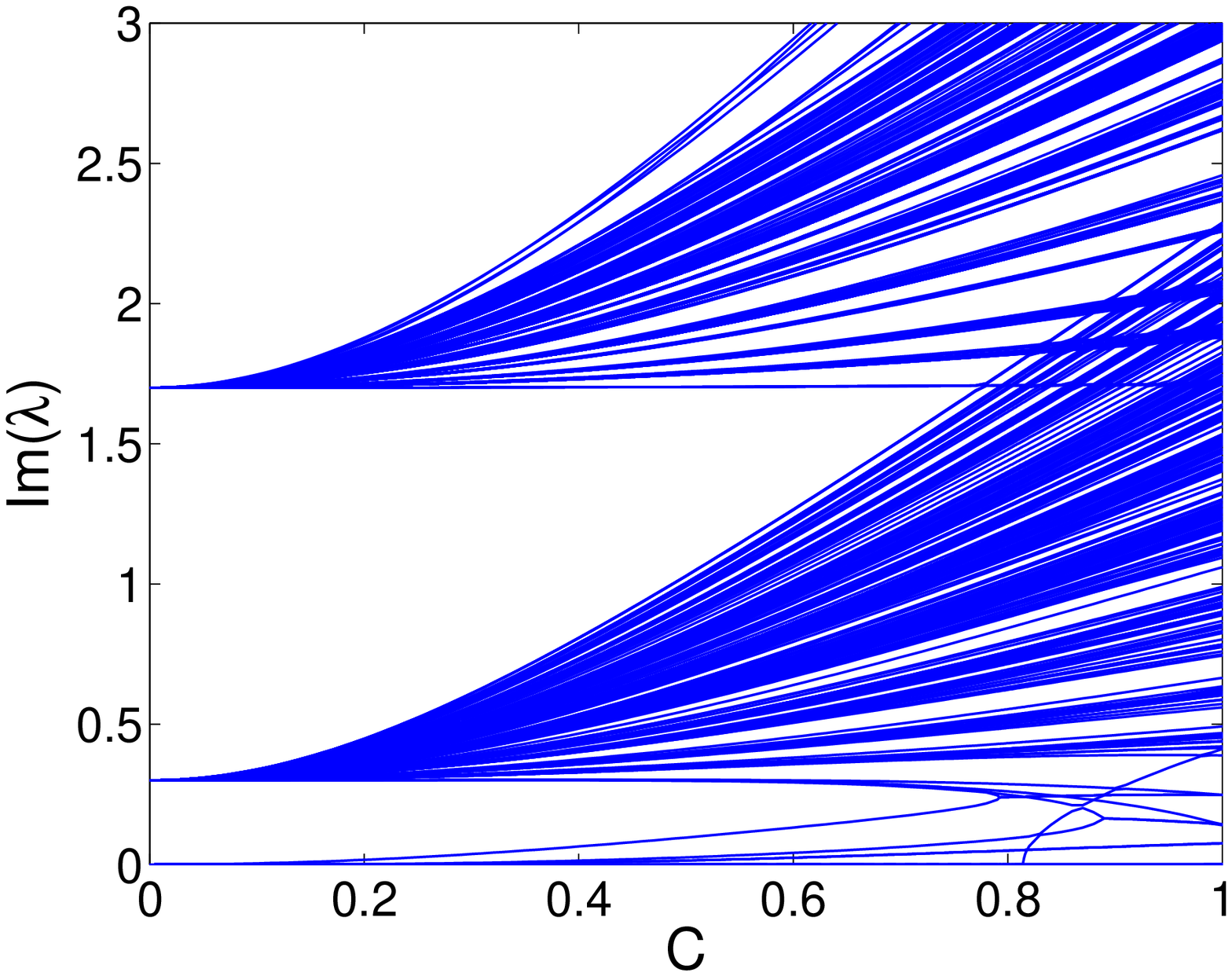} &
\includegraphics[width=6cm]{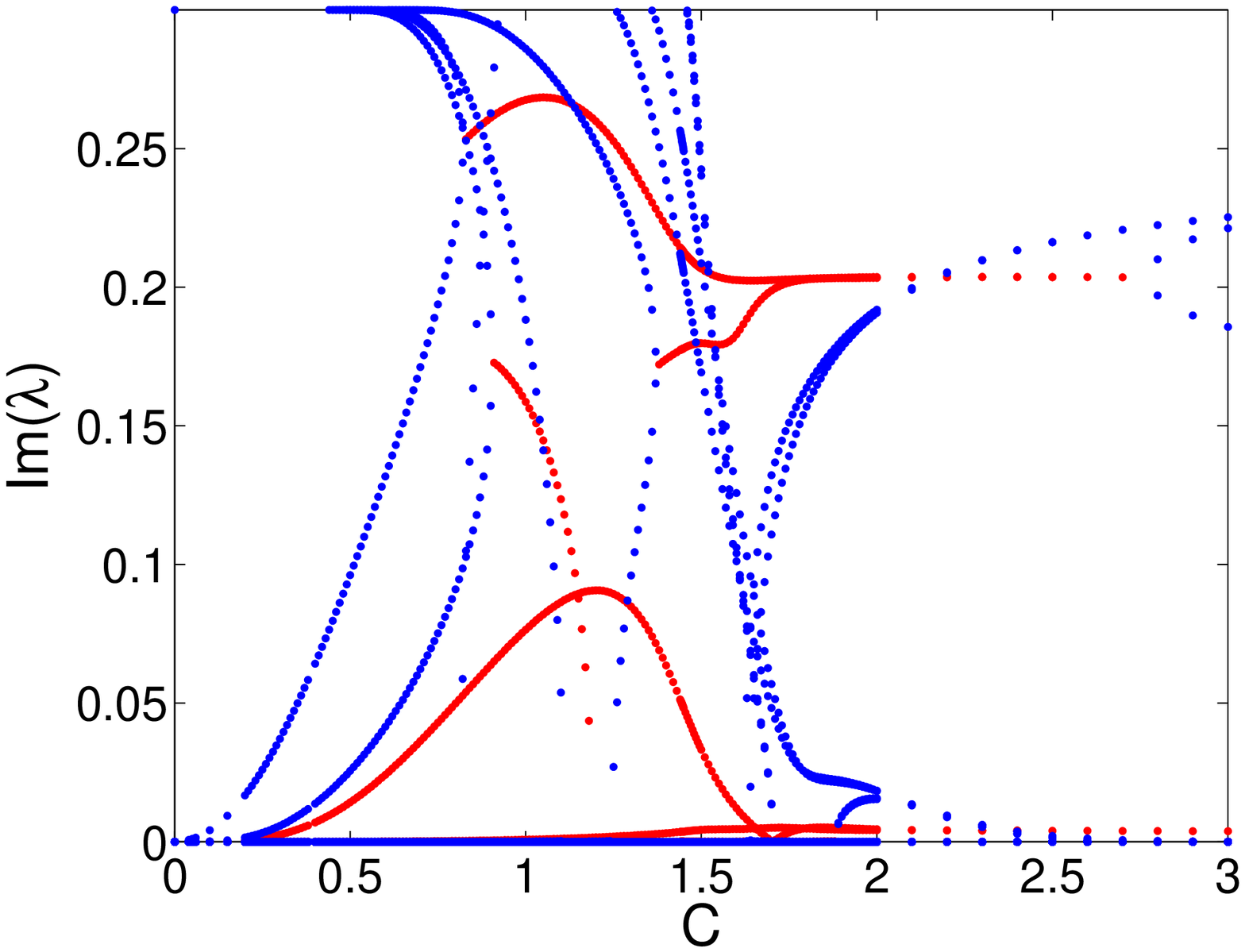} &
\includegraphics[width=6cm]{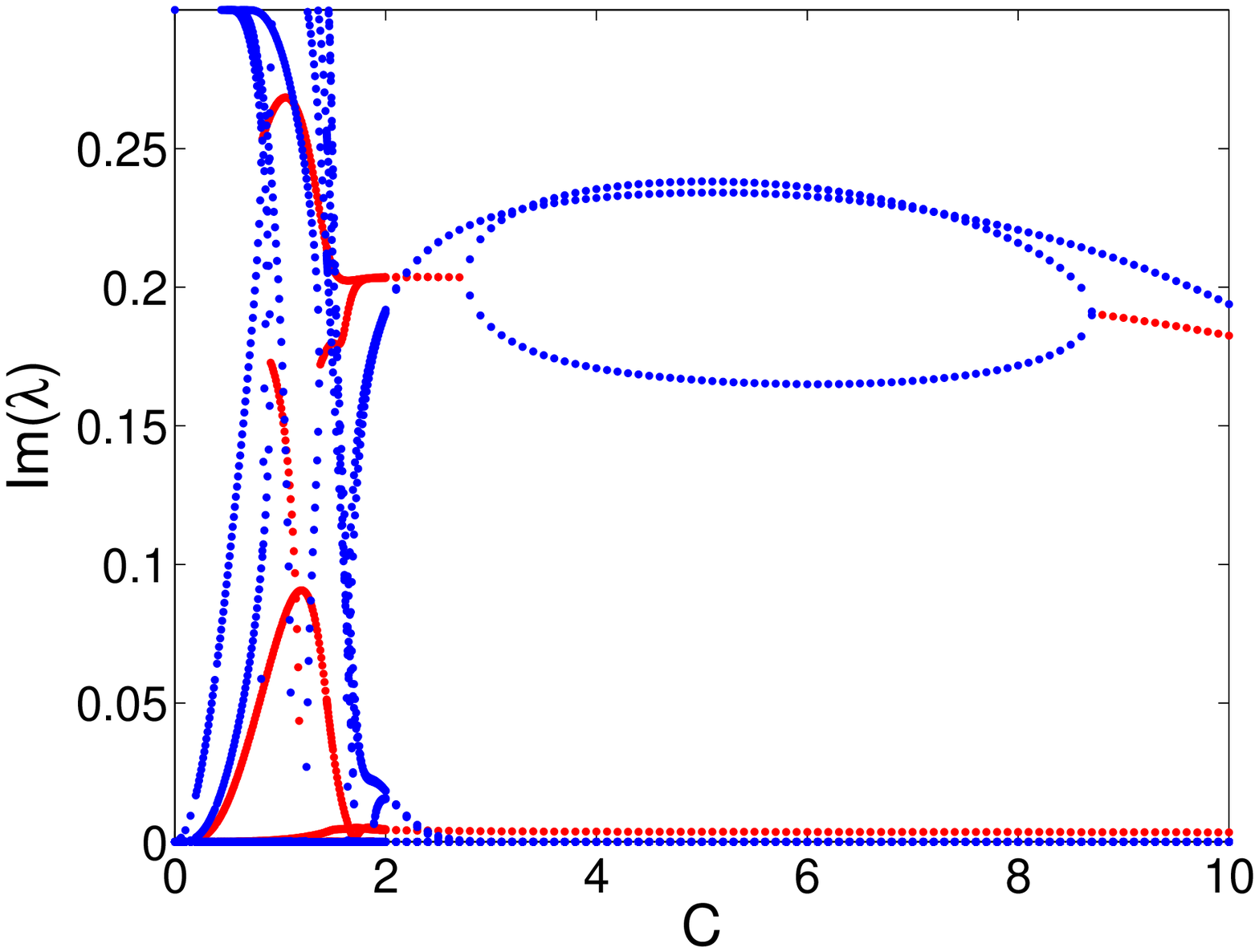} \\
\includegraphics[width=6cm]{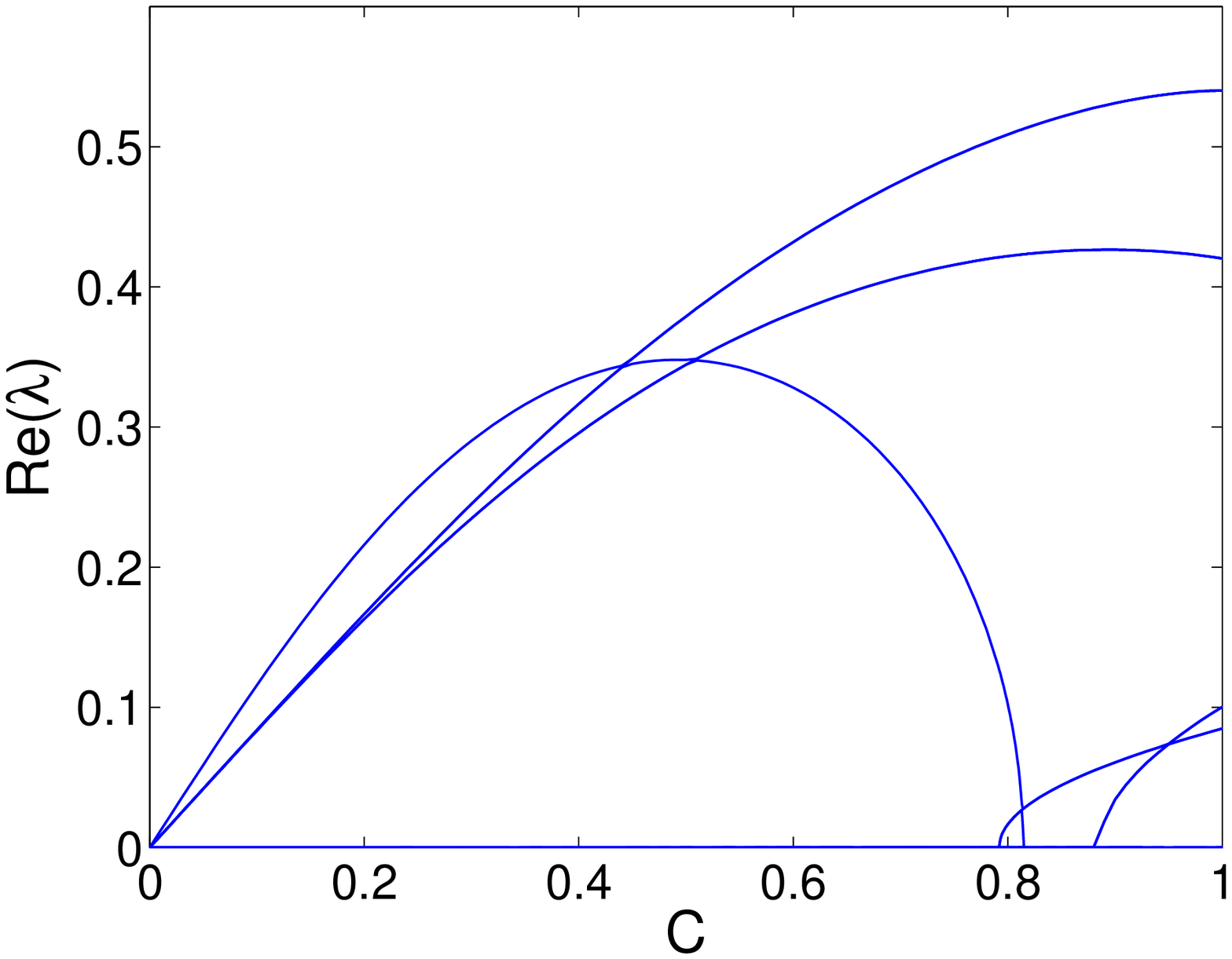} &
\includegraphics[width=6cm]{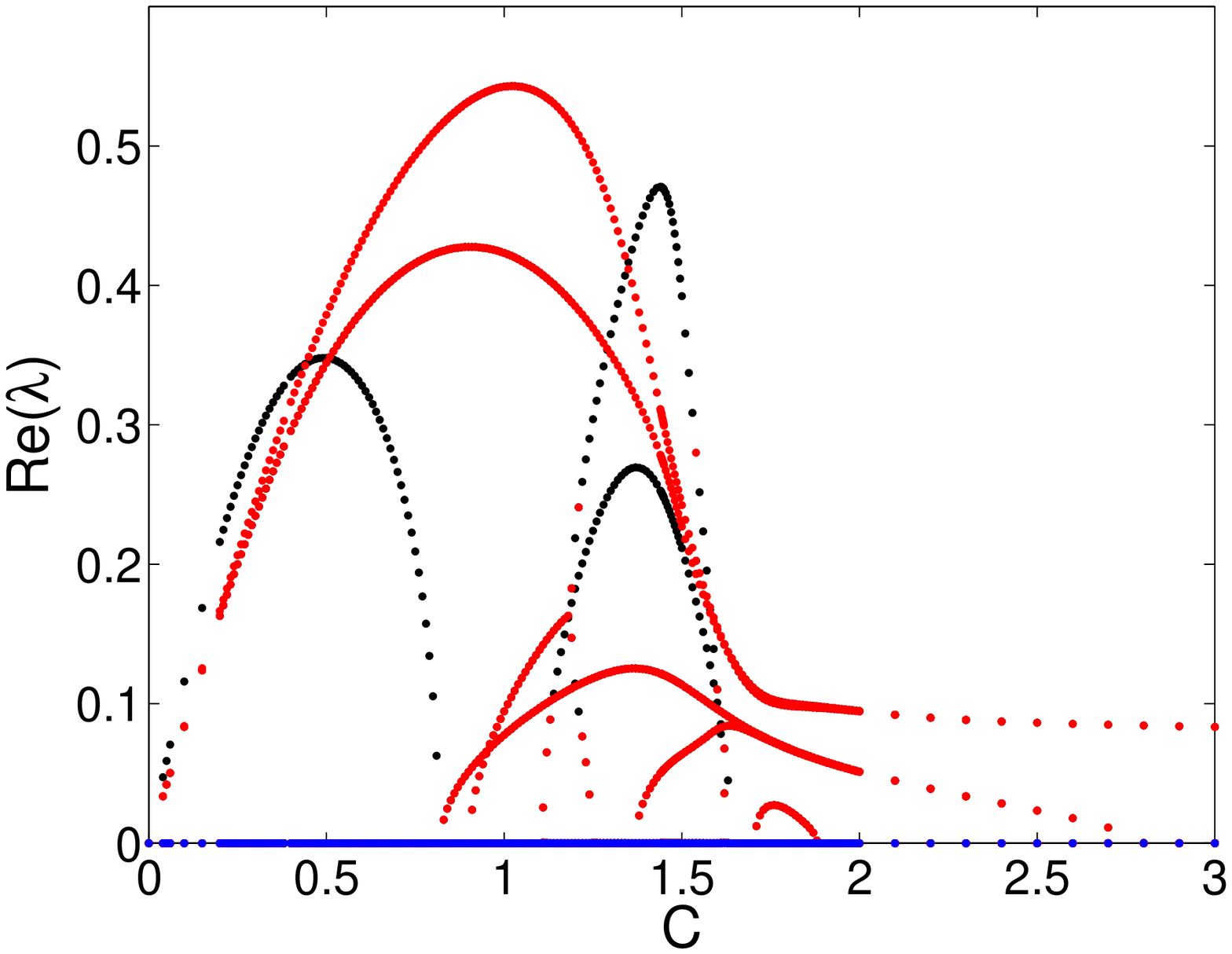} &
\includegraphics[width=6cm]{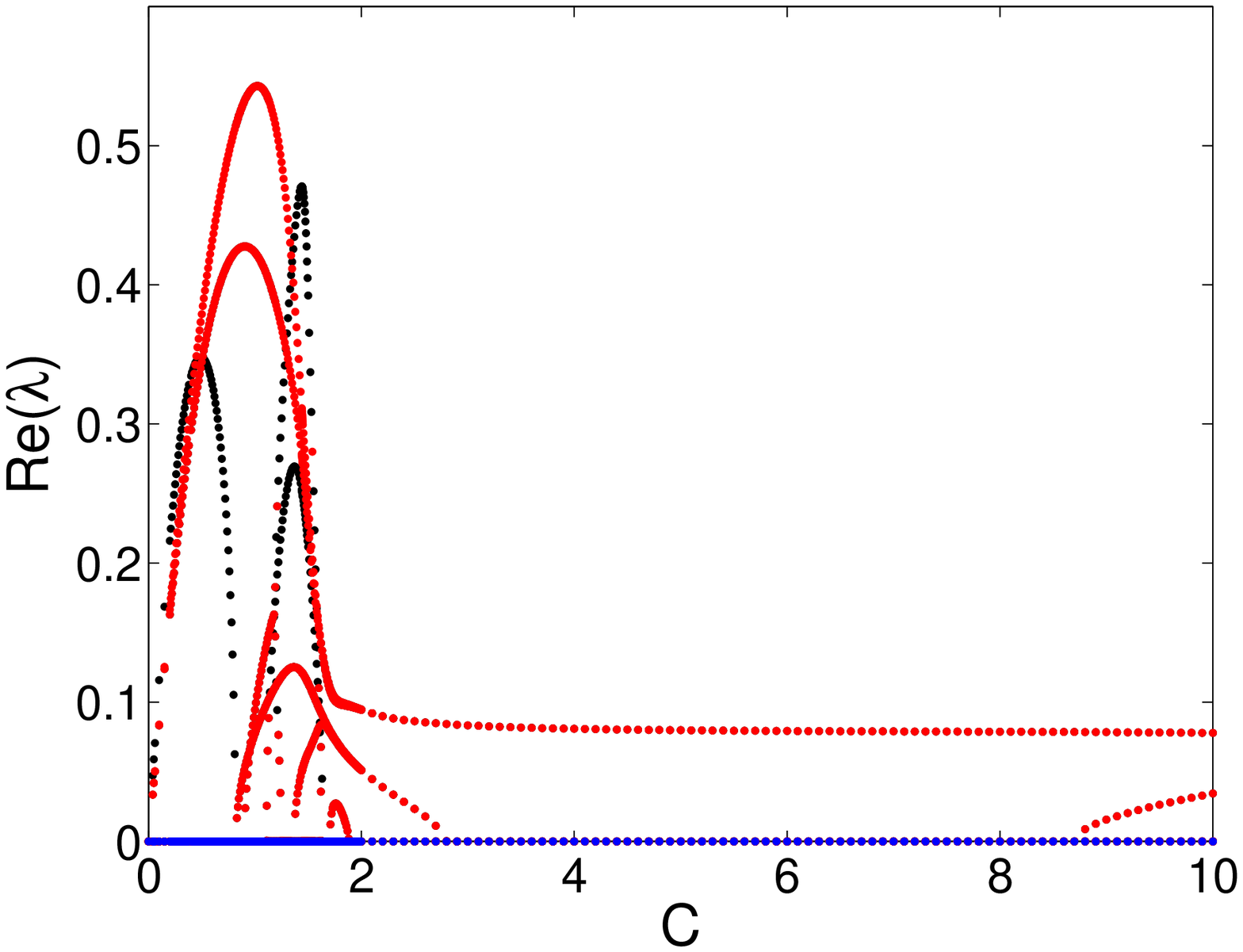} \\
\end{tabular}
\caption{Dependence of the stability eigenvalues for the 9-site soliton with respect to $C$ for $\omega=0.7$. The left panels show the spectrum for a     lattice with $N=20$ sites in each dimension. As the system is too small, continuation for couplings higher than $C=1$ is not pursued. In addition, computation of the full spectrum for large lattices is not possible. For this reason, the middle and right panels show a partial spectrum (i.e., the imaginary and real part of the eigenvalues with the lowest imaginary part) up to $C=3$ and $C=10$, respectively for $N=200$. We also indicate with red (black) dots, the eigenvalues corresponding to oscillatory (exponential) instabilities, whereas blue dots correspond to stable eigenmodes. In each case, the top panels show the imaginary parts and the bottom panels the real parts (corresponding to unstable growth) of the relevant eigenvalues.}
\label{fig:stab9site}
\end{figure}

Among all the solutions in the AC limit, remarkably we find that the one that leads to the $S=0$ solitary waves of the continuum limit -- discussed in the previous section -- is the 9-site soliton, i.e. $u_{0,0}=u_{0,\pm1}=u_{\pm1,0}=u_{\pm1,\pm1}=\sqrt{1-\omega}$. The $v$ field must be vanishing at $C=0$, as can be seen from Eq.~(\ref{eq:stat}). The left panel of Fig.~\ref{fig:profs} shows the profile of a typical such solution at finite coupling. Notice that in the left panels of Fig. \ref{fig:profs}, one can observe that, contrary to the soliton in the continuum, the imaginary part of $u_{n,m}$ is not null; however, $||\mathrm{Im}(u_{n,m})||_\infty$ tends asymptotically to 0 when $C\rightarrow\infty$. An additional advantage of the AC limit is that the decoupled nature of the lattice enables us to analytically calculate the spectrum at $C=0$. This consists of 9 pairs of eigenvalues at 0, $N^2-9$ pairs at $\lambda=\pm i(1+\omega)$ and $N^2$ pairs at $\lambda=\pm i(1-\omega)$. As in the 1D case, some of the eigenvalues become real when the coupling is switched on and the soliton is unstable for finite coupling. In particular, 7 eigenvalue pairs detach from zero whereas 2 additional pairs remain at zero for every coupling. Of the seven remaining pairs, two become imaginary and five pairs acquire a real part; for very low coupling $C\lesssim0.01$, four among these real pairs experience Hamiltonian Hopf bifurcations yielding complex quartets. This scenario persists for $C<0.80$. Beyond this value, a rather complicated scenario arises, as can be seen in Fig. \ref{fig:stab9site}, where the dependence of the stability eigenvalues as a function of the coupling constant $C$ for $\omega=0.7$ is shown. One can observe that for large $C$ there are only two sources of oscillatory instabilities: (1) for $C\gtrsim2.7$, one of the quartets that exists for small $C$ persists even for high $C$; (2) at $C\approx8.8$, an additional Hamiltonian Hopf bifurcation takes place. Contrary to the Soler model \cite{JPA}, the $2i\omega$ eigenvalue is not present either in the discrete or in the continuum limit.

\begin{figure}
\begin{tabular}{cc}
\includegraphics[width=6cm]{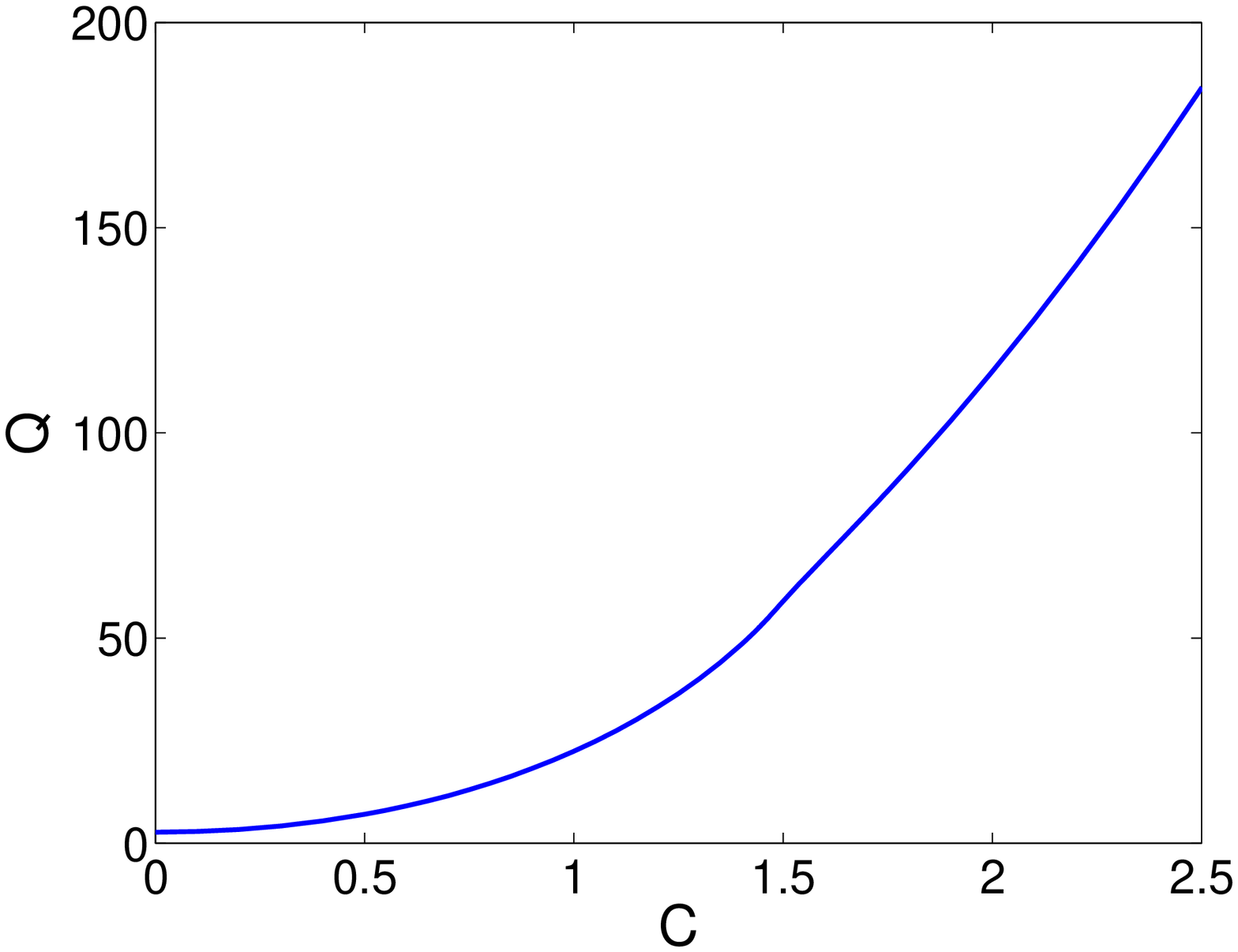} &
\includegraphics[width=6cm]{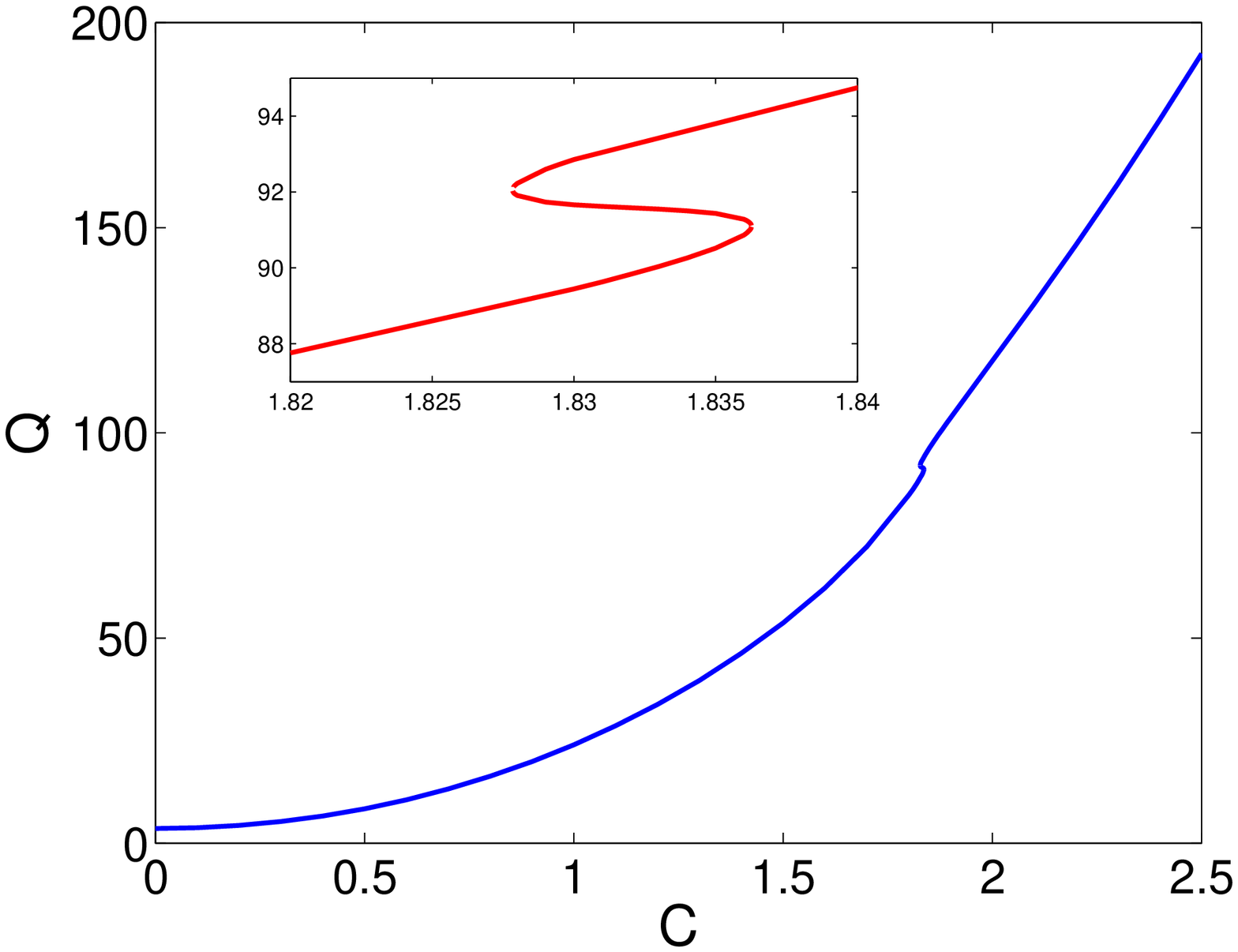} \\
\end{tabular}
\caption{Dependence of the charge (power) $Q$ on the coupling constant for the 9-site solitons with $\omega=0.7$ (left) and $\omega=0.6$ (right). The  inset in the latter zooms in the region where three branches coexist for a narrow interval of values of $C$.}
\label{fig:charge9site}
\end{figure}

It is intriguing to note that for the discretization considered
at least one of the Hamiltonian Hopf bifurcations persists for
large values of $C$. Such a feature has been previously
encountered in the finite difference analysis of the Soler model
in~\cite{JPA}. Another interesting feature that occurs below a certain critical frequency is the existence of two branches of solitons, one starting at $C=0$ and another one finishing at $C\rightarrow\infty$. These branches are connected by an intermediate one through two saddle-center bifurcations i.e., the dependence of $Q$ on $C$ has an S-shaped form. This type of bifurcation point arises for higher values of $C$ when $\omega$ decreases.
Figure \ref{fig:charge9site} shows the relevant dependence of the charge with respect to $C$ for $\omega=0.7$ and $\omega=0.6$. We can observe in the latter case the existence of this intermediate branch, as well as the associated fold
points.

In a similar fashion to the 1D case~\cite{JPA}, 1-site solitons also exist and can be continued to the continuum limit. In the AC limit, they are given by $u_{0,0}=\sqrt{1-\omega}$, while $v_{0,0}=0$. The spectrum at $C=0$ comprises $N^2-1$ pairs at $\lambda=\pm i(1+\omega)$, $N^2$ pairs at $\lambda=\pm i(1-\omega)$ and a sole pair at $\lambda=0$. As the single pair at 0 must remain there because of the U(1) symmetry, no eigenvalue can depart from 0 and the soliton is stable, at least for small values of the coupling parameter $C$.

\begin{figure}
\begin{tabular}{cc}
\includegraphics[width=6cm]{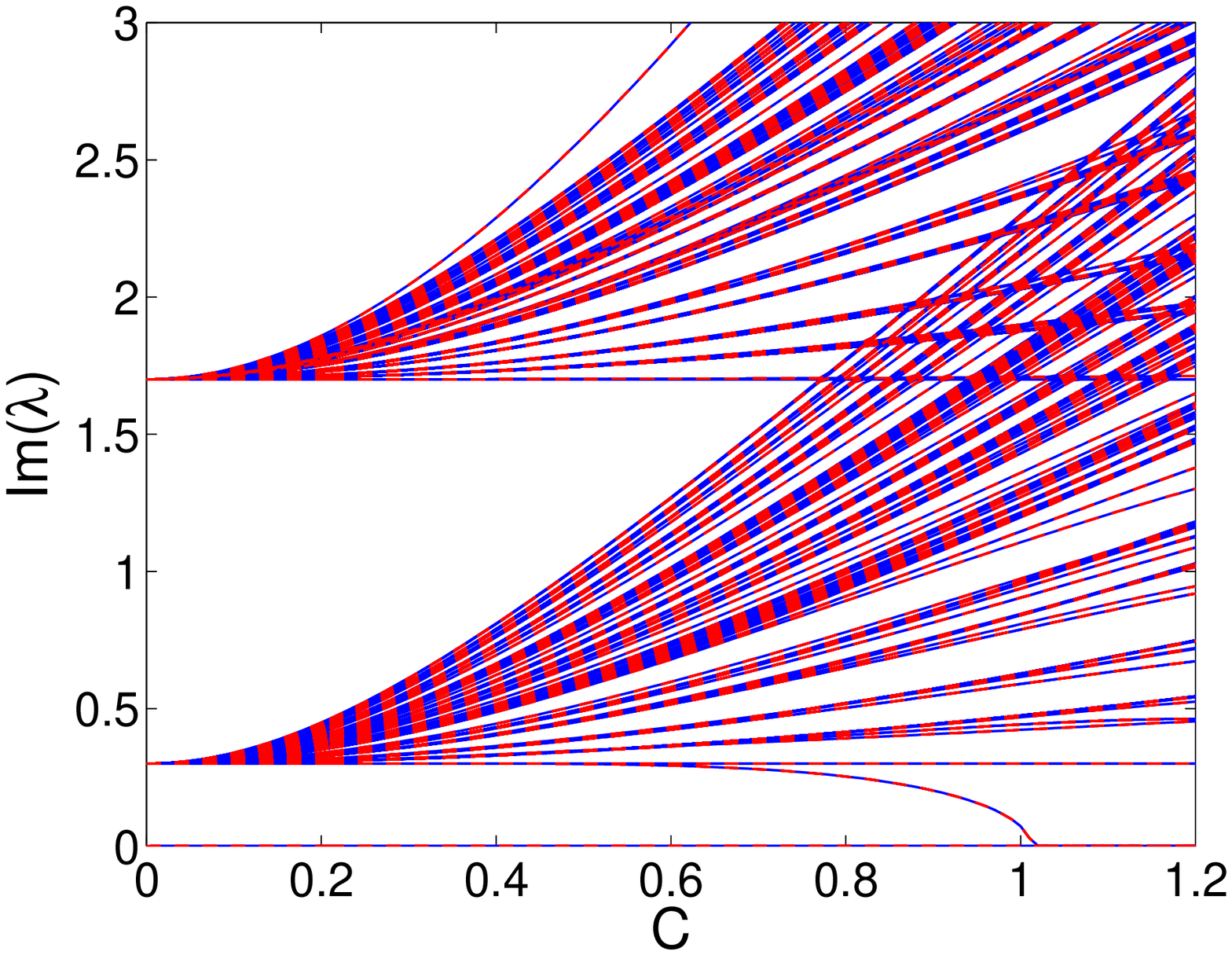} &
\includegraphics[width=6cm]{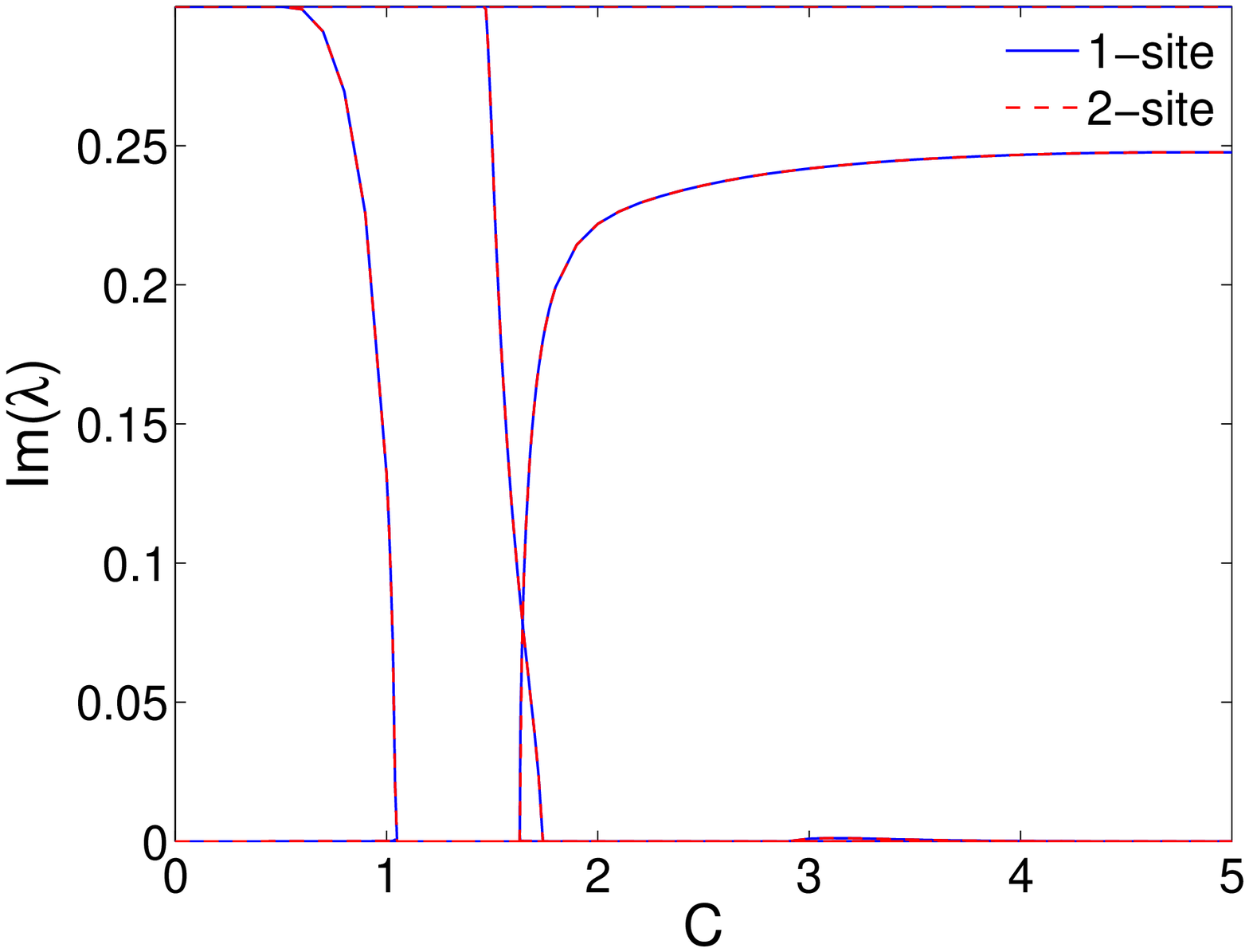} \\
\includegraphics[width=6cm]{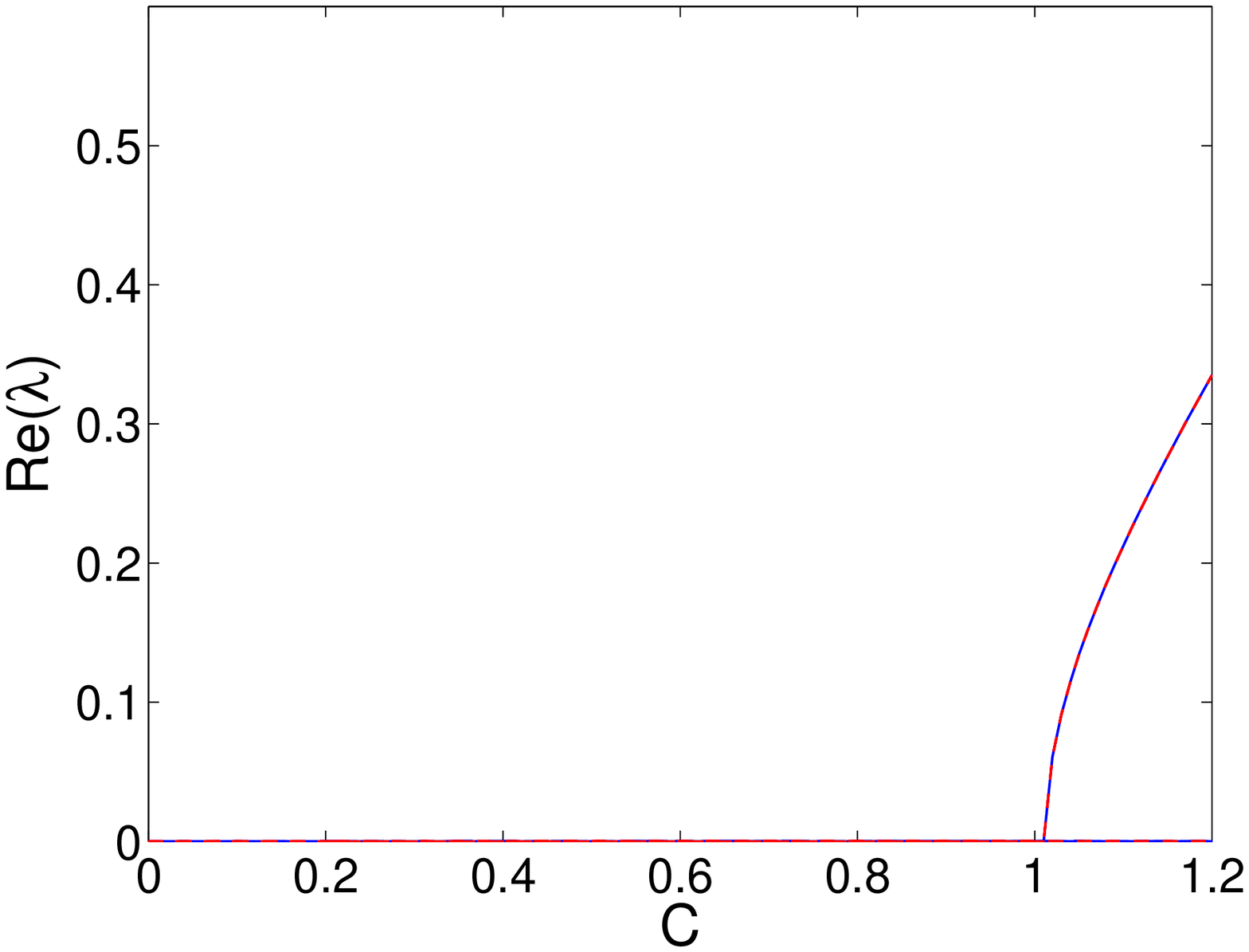} &
\includegraphics[width=6cm]{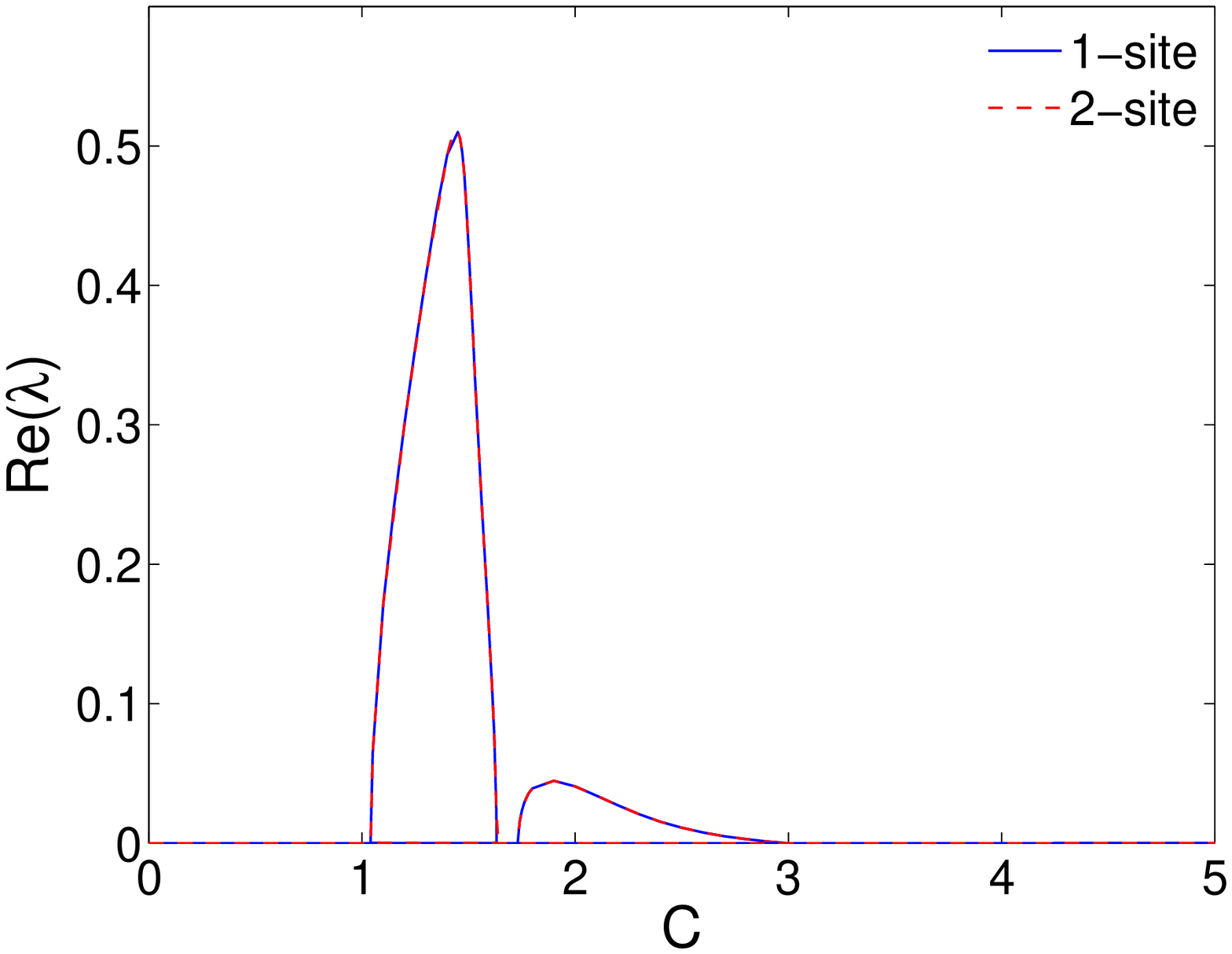}
\end{tabular}
\caption{Dependence of the stability eigenvalues for the 1-site soliton (full blue line) and the 2-site soliton (dashed red line) with respect to $C$ for $\omega=0.7$. The left panels show the spectrum for a lattice with $N=20$ sites in each dimension. The right panels show the relevant eigenvalues [i.e.,
only part of the spectrum as concerns especially $\mathrm{Im}(\lambda)$] for $N=200$. Notice that the two spectra are almost identical and, consequently, full blue and dashed red lines match almost perfectly. In every case, the top panels show the imaginary parts and the bottom panels the real parts (corresponding to unstable growth) of the relevant eigenvalues.}
\label{fig:stab12site}
\end{figure}

The middle panel of Fig.~\ref{fig:profs} shows the profile of a 1-site soliton for $\omega=0.7$ and $C=1$. Notice that such solitons, for every coupling, possess the following properties:

\begin{itemize}

\item $\Re(u_{n,m})=0$ if $n$ and $m$ are odd,
\item $\Im(u_{n,m})=0$ if $n$ and $m$ are even,
\item $\Re(v_{n,m})=0$ if $n$ is odd and $m$ is even,
\item $\Im(v_{n,m})=0$ if $n$ is even and $m$ is odd,

\end{itemize}

\noindent
resembling the properties of the soliton of Fig. 2 of \cite{Tran} in the large $C$ limit. This also endows the solitons' real and imaginary parts with a
``staggered'' structure with alternating rows missing, as seen in the middle panels of Fig.~\ref{fig:profs}. Figure \ref{fig:stab12site} shows the dependence of the complex eigenvalues on $C$. One can see that the soliton is stable (see the real part in the bottom left panel of the figure)
for $C$ below $1.04$. At this critical point there is a bifurcation caused by a mode that destabilizes, after bifurcating from the linear modes band. Above this point, the soliton is exponentially unstable, becoming stable again at $C=1.631$. However, at $C=1.73$ it experiences a similar instability anew, while the structure finally stabilizes  $\forall C\geq3.0$ in this case of $\omega=0.7$.

Another interesting structure is the 2-site soliton, which, in the continuum limit is reminiscent of the soliton of Fig. 5 in \cite{Tran}.  In the AC limit, such a wave structure is given by $u_{0,0}=u_{0,1}=\sqrt{1-\omega}$, while once again the $v$ field is vanishing. The spectrum at $C=0$ is composed by $N^2-2$ pairs at $\lambda=\pm i(1+\omega)$, $N^2$ pairs at $\lambda=\pm i(1-\omega)$ and two pairs at $\lambda=0$. As in the 9-site case, the two pairs remain invariant at $0$, and thus no eigenvalue departs from 0, making the structure spectrally stable for small $C$.

The right panel of Fig.~\ref{fig:profs} shows the profile of a 2-site soliton for $\omega=0.7$ and $C=1$. Here too, similarly to the 1-site soliton, we observe a staggered pattern with the following features:

\begin{itemize}

\item $\Re(u_{n,m})=0$ if $n$ is odd,
\item $\Im(u_{n,m})=0$ if $n$ is even,
\item $\Re(v_{n,m})=0$ if $n$ is even,
\item $\Im(v_{n,m})=0$ if $n$ is odd.

\end{itemize}

Figure \ref{fig:stab12site} shows the dependence of the linearization eigenvalues with $C$. One can see that the behavior is essentially the same as in the case of the 1-site soliton, with {the bifurcations taking place at the same points as in the 1-site case. In fact, the spectrum is almost identical in both 1-site and 2-site case, as the panels of Fig.~\ref{fig:stab12site} show}.

\begin{figure}
\begin{tabular}{ccc}
\includegraphics[width=6cm]{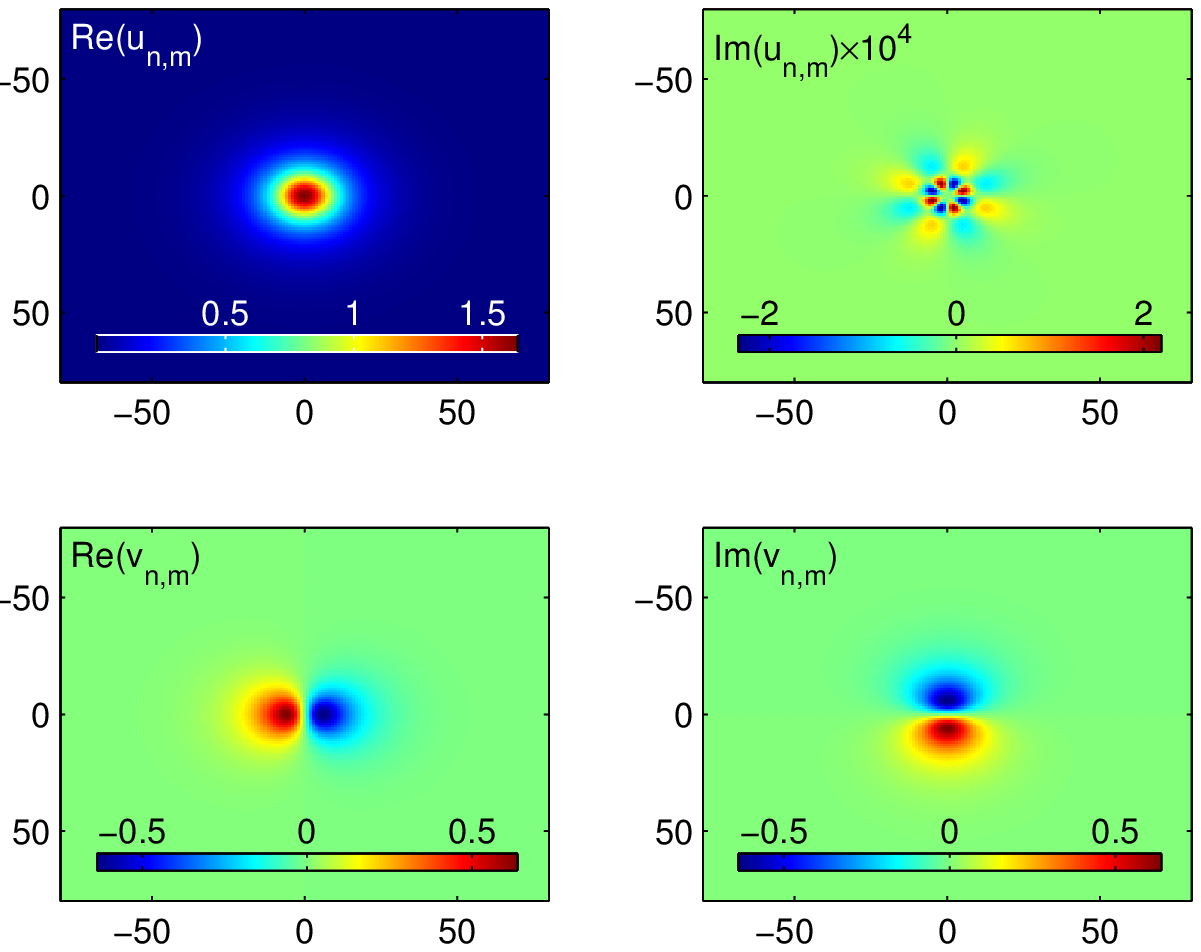} &
\includegraphics[width=6cm]{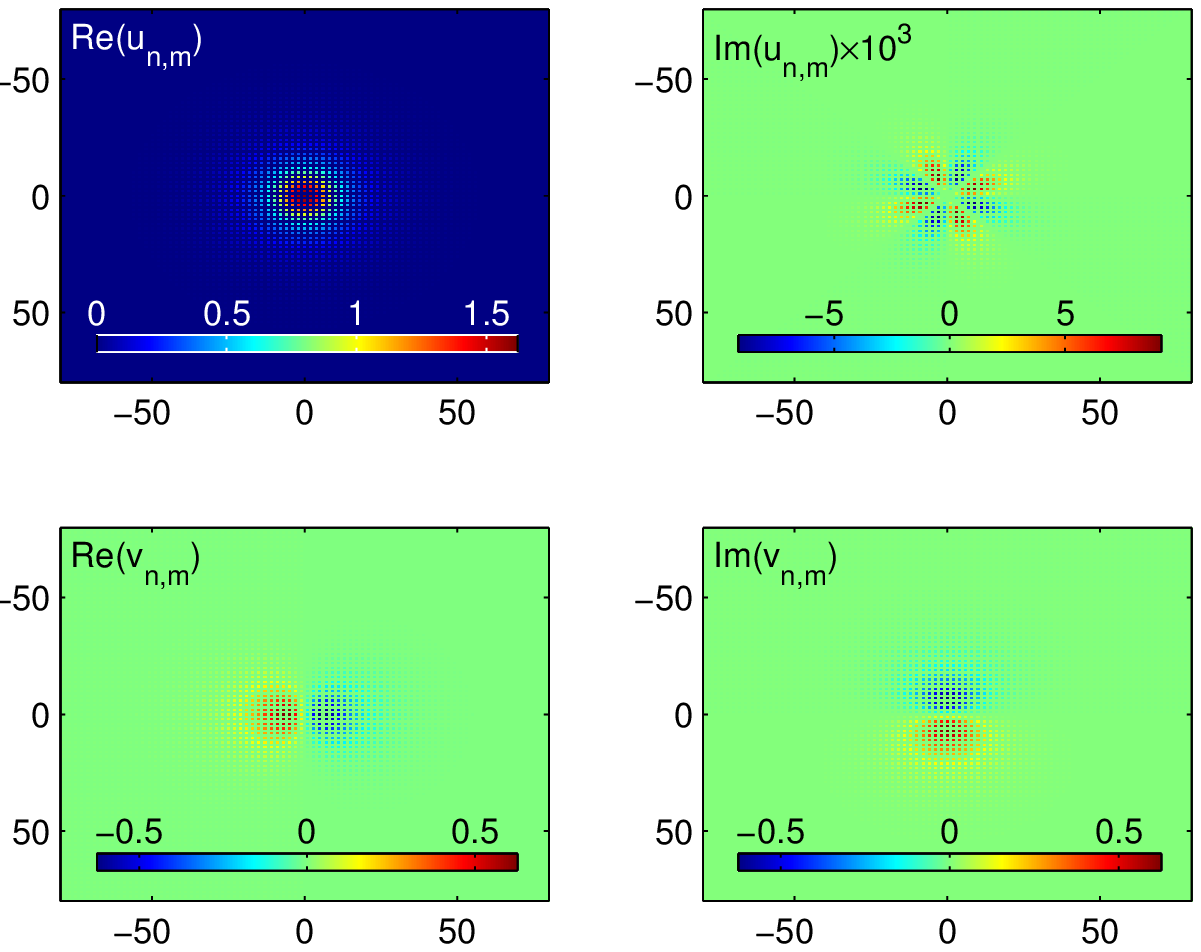} &
\includegraphics[width=6cm]{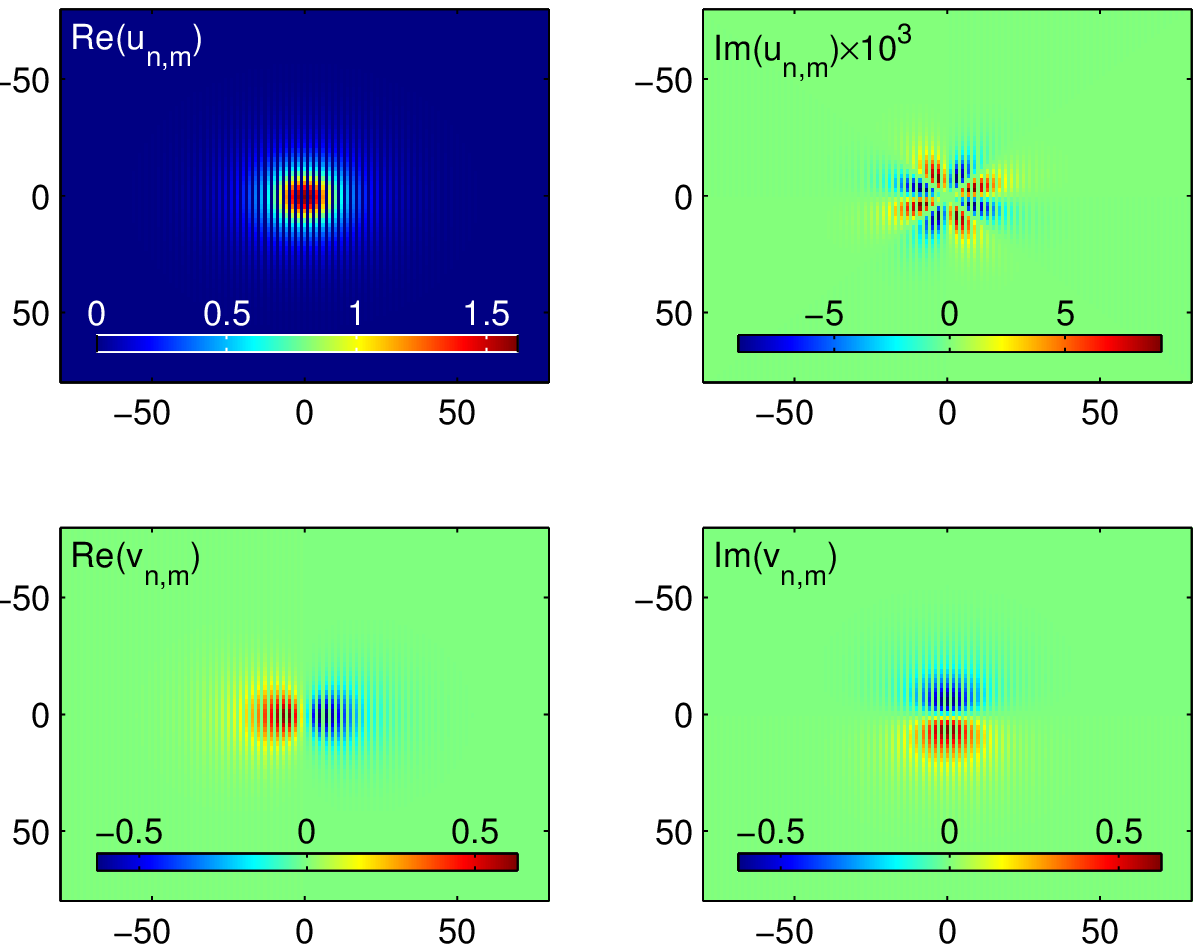} \\
\end{tabular}
\caption{Profiles of the 9-site (left set of $2 \times 2$ panels), 1-site (middle set of panels) and 2-site (right set of panels) soliton with $\omega=0.5$ and $C=5$. We can clearly observe in the 9-site case that the configuration approaches the continuum soliton. In the 1-site and 2-site cases, the staggered patterns with the previously mentioned characteristics persist.}
\label{fig:profscont}
\end{figure}

We have also considered the stability of the above mentioned configurations for fixed coupling near the continuum limit and variable frequency. Figure \ref{fig:profscont} shows the profile of such solutions for $\omega=0.5$ and $C=5$.

\begin{figure}
\begin{tabular}{cc}
\includegraphics[width=6cm]{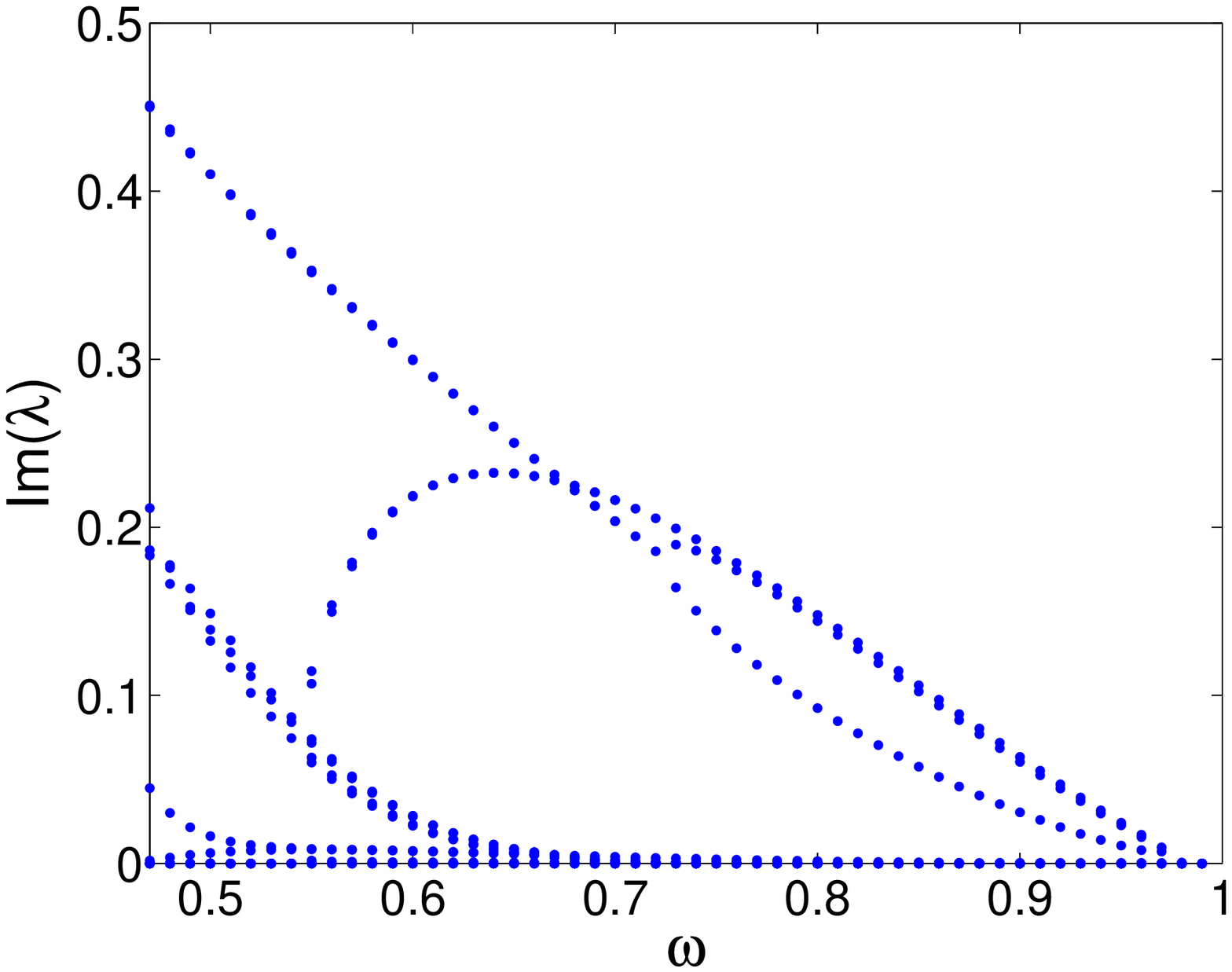} &
\includegraphics[width=6cm]{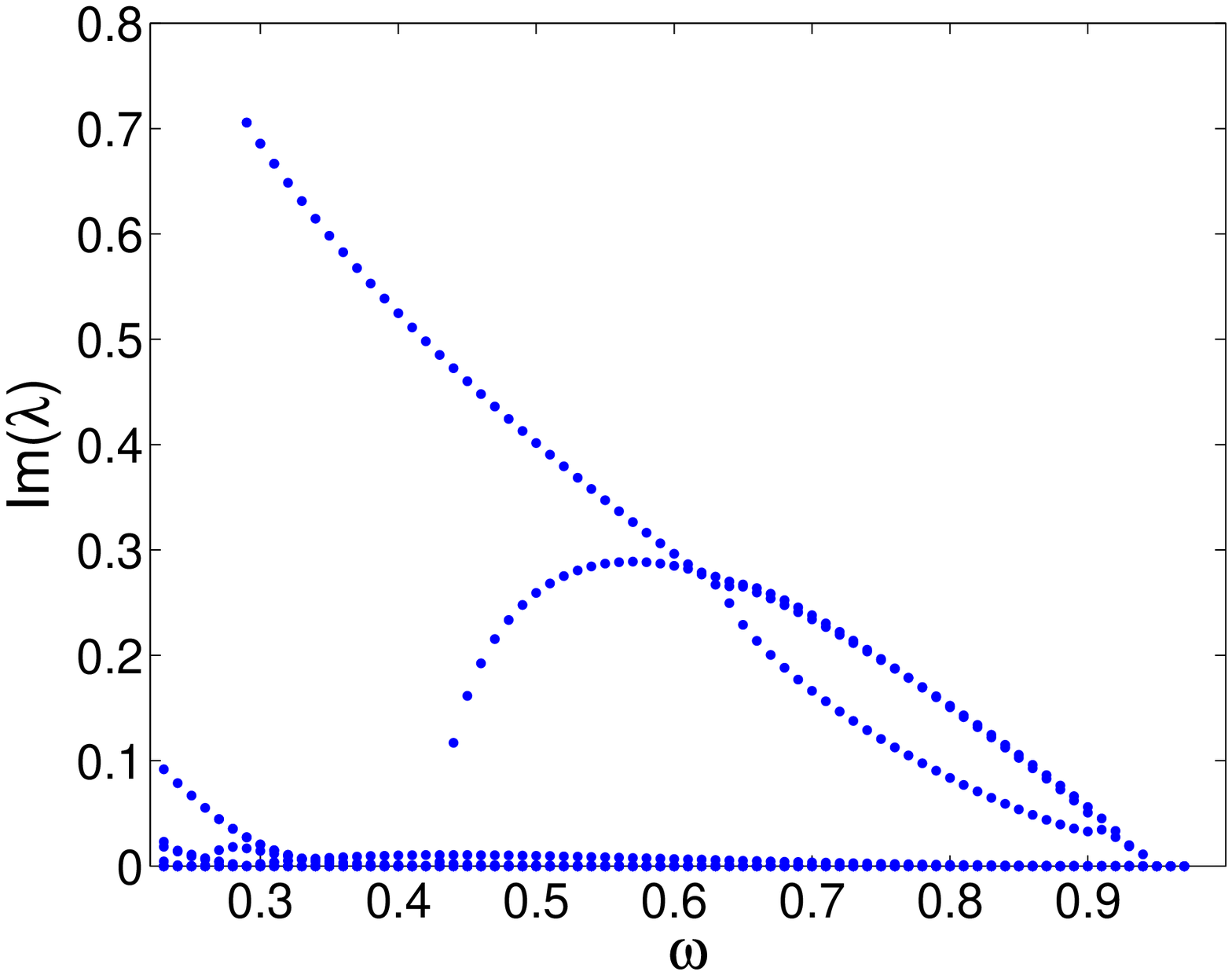} \\
\includegraphics[width=6cm]{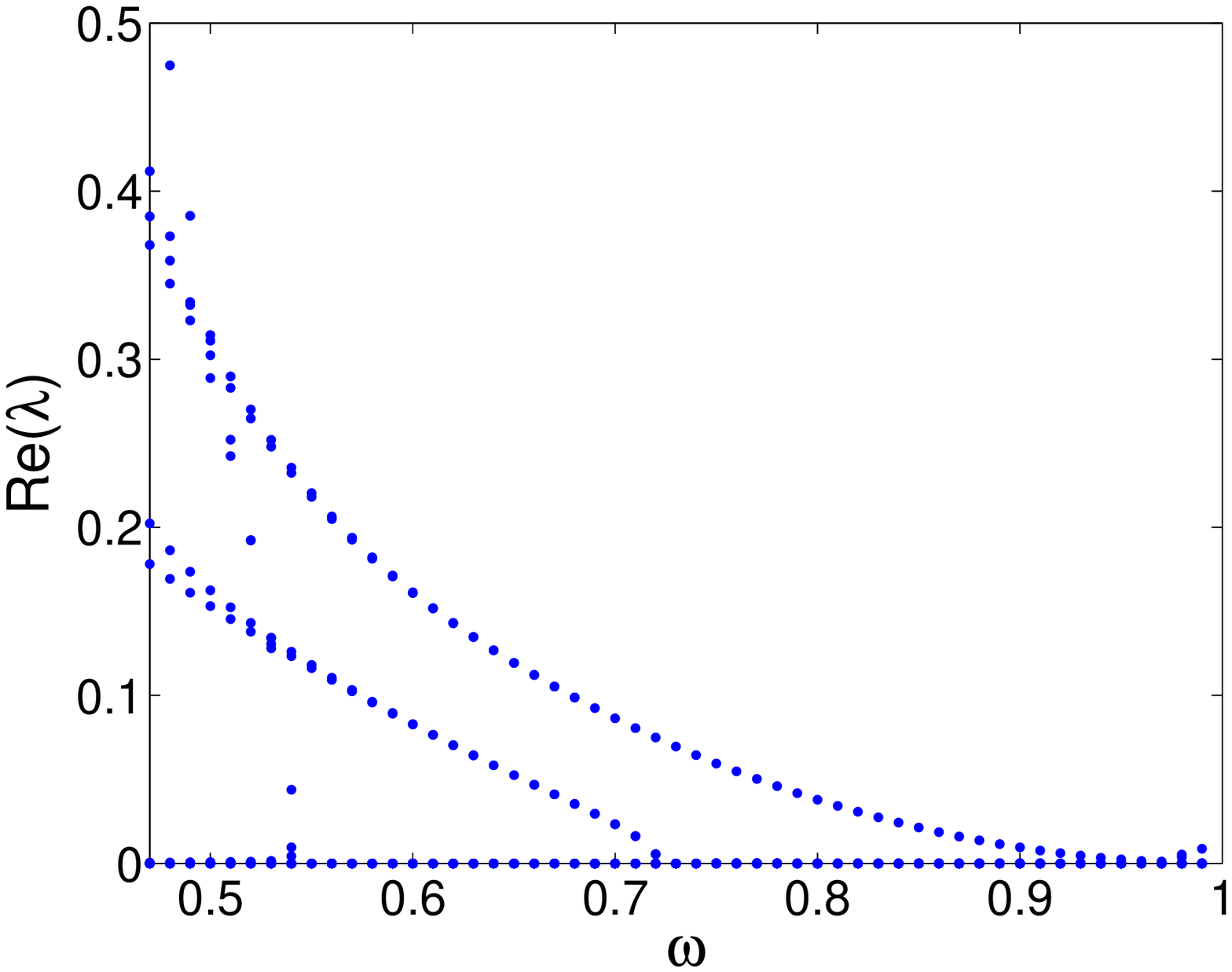} &
\includegraphics[width=6cm]{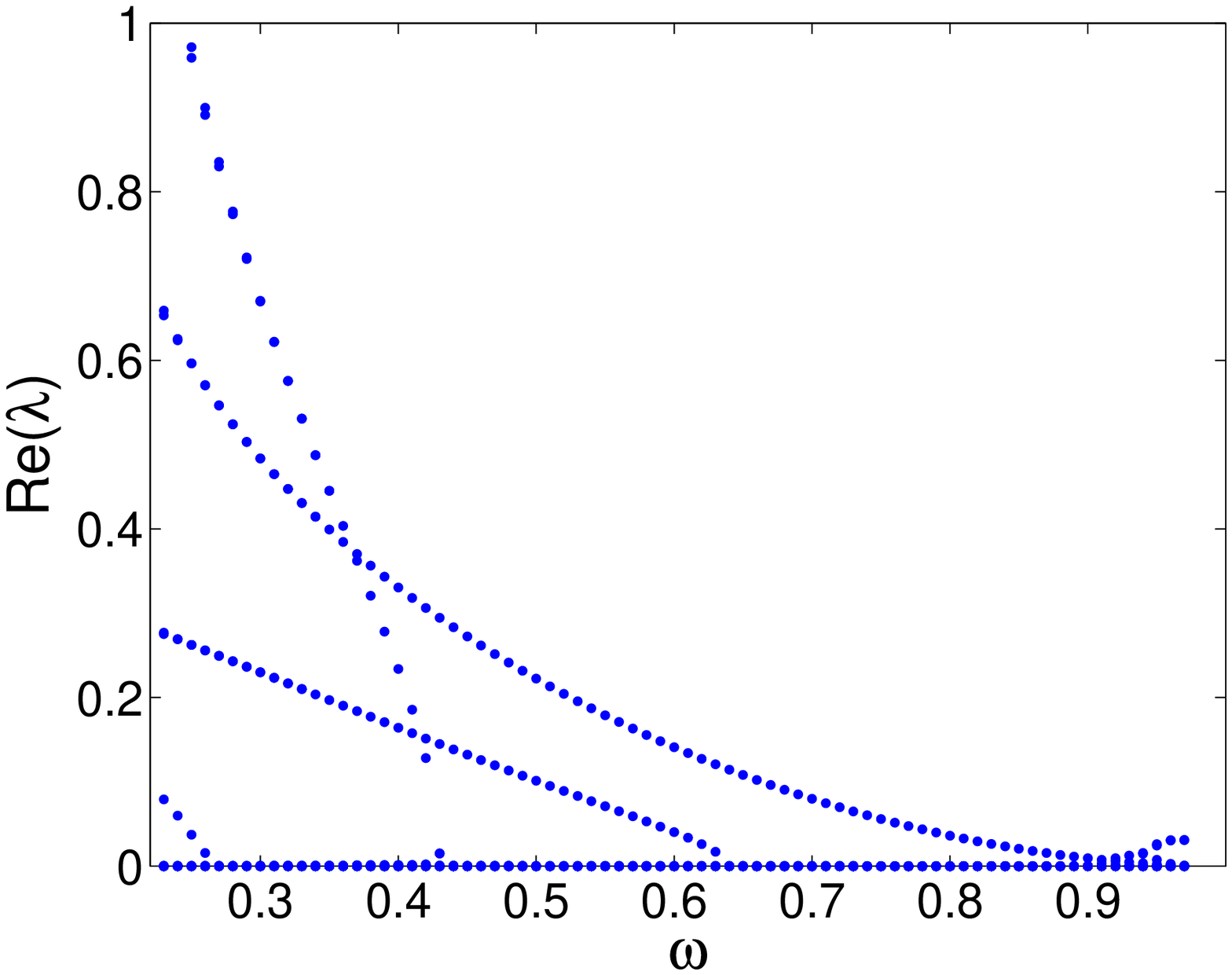}
\end{tabular}
\caption{{Dependence of the relevant stability eigenvalues for the 9-site soliton with respect to $\omega$ for $C=2.5$ (left panels) and $C=5$ (right panels); $N=200$ in both cases. The bottom panels suggest that for both values the soliton is dynamically unstable in nearly the full interval of frequencies between $\omega=0$ and $\omega=1$. Notice the different scales on the $\omega$-axis in the left and right panels.}}
\label{fig:stab9sitecont}
\end{figure}

\begin{figure}
\begin{tabular}{cc}
\includegraphics[width=6cm]{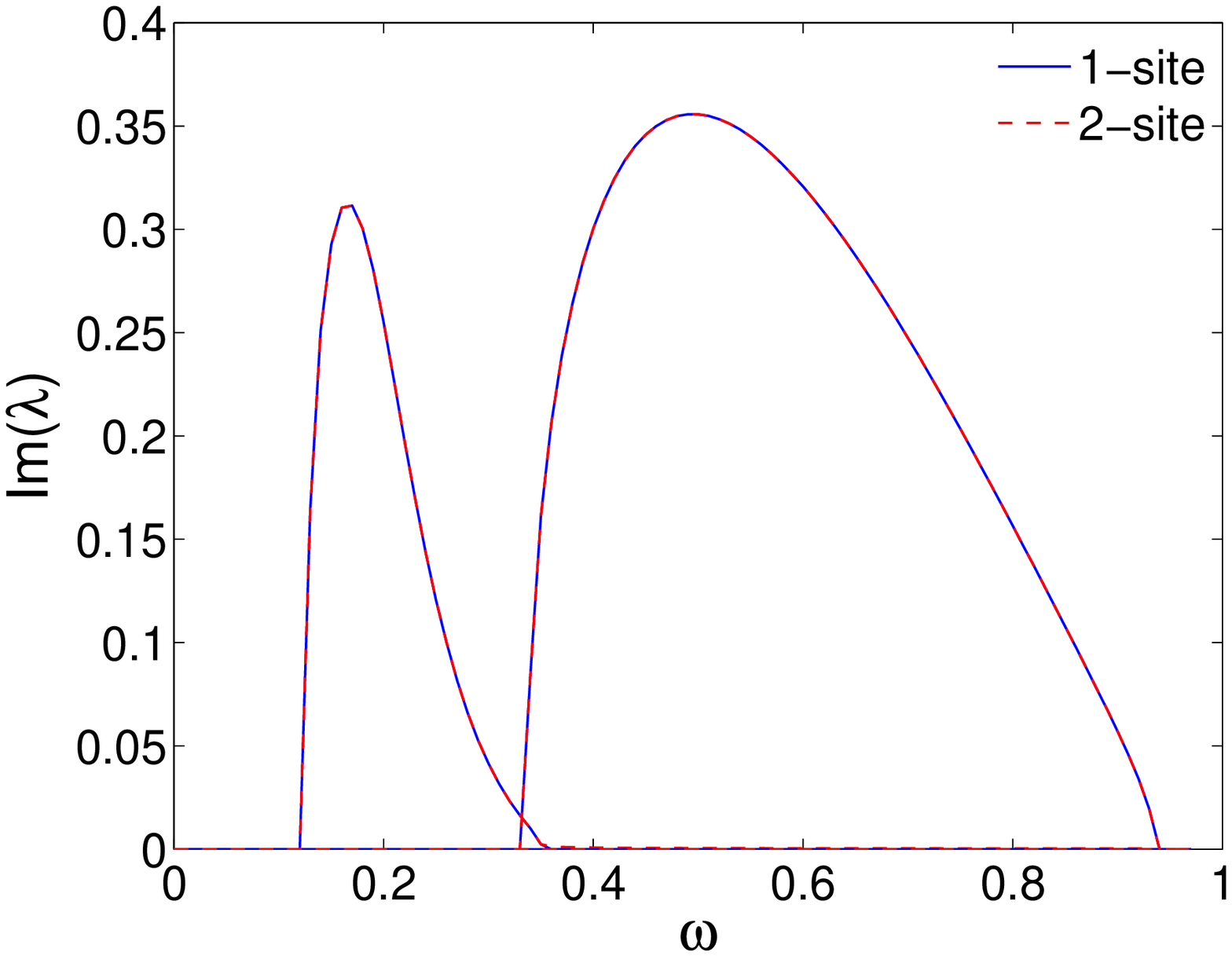} &
\includegraphics[width=6cm]{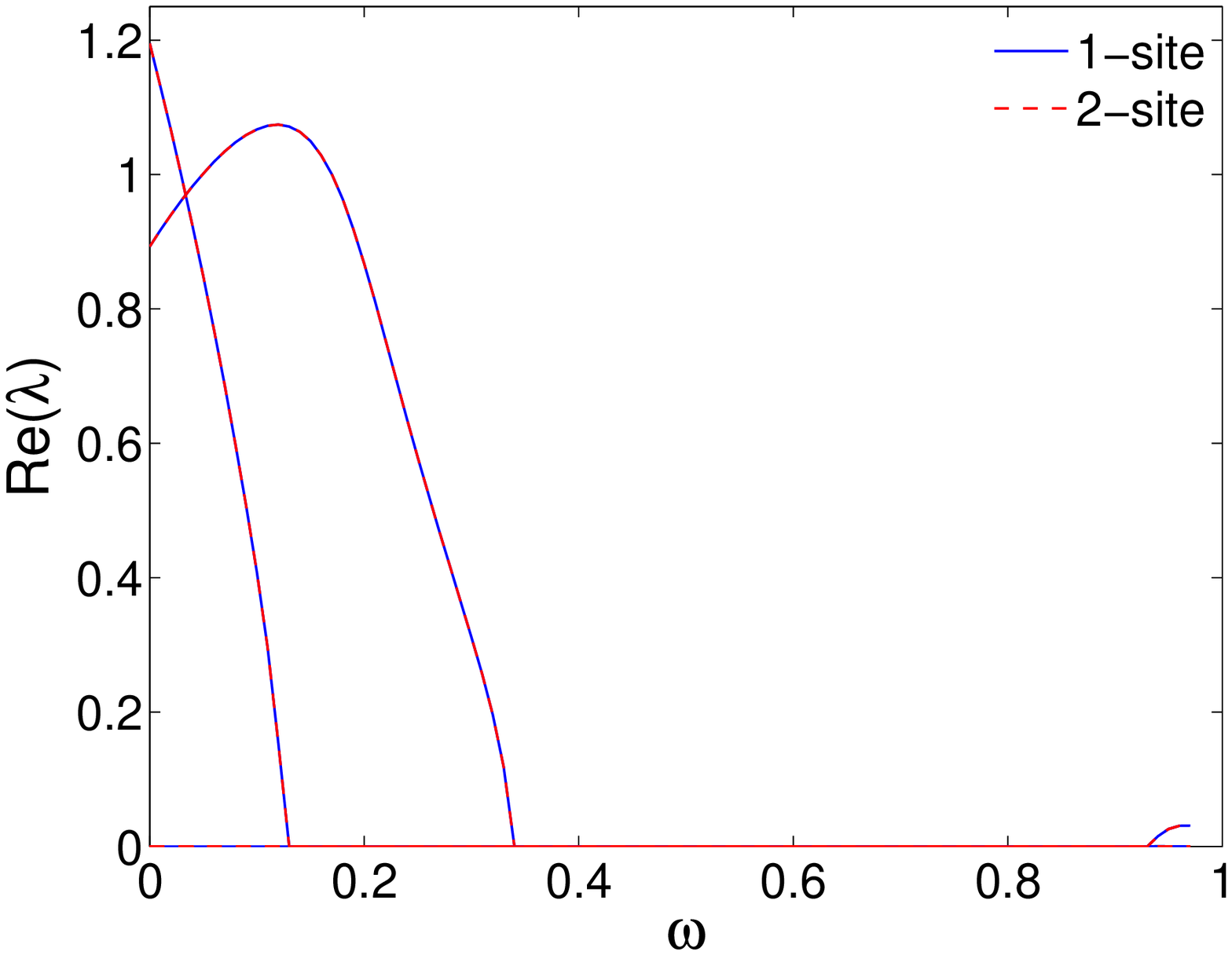} \\
\end{tabular}
\caption{Dependence of the relevant stability eigenvalues for the 1-site (full blue line) and 2-site (red dashed line) solitons with respect to $\omega$ for $C=5$. $N=200$ in both cases. Only a partial spectrum is shown. Notice that the two spectra are almost identical and, consequently, full blue and dashed red lines match almost perfectly.}
\label{fig:stab12sitecont}
\end{figure}

Figure \ref{fig:stab9sitecont} shows, for $C=2.5$ and $C=5$, the dependence of the complex eigenvalues with $\omega$ for the 9-site solitons. Notice that solitons only exist above a critical value of $\omega(C)$; this critical value decreases with $C$. For instance, solitons only exist for $\omega\geq0.47$ ($\omega\geq0.22$) if $C=2.5$ ($C=5$). This is caused by a bifurcation similar to that shown in Fig. \ref{fig:charge9site}. This bifurcation is not found (at least for the considered coupling $C=5$) in the 1-site and 2-site solitons (see Fig. \ref{fig:stab12sitecont}). In the latter case, we observe that while generally dynamical instabilities may arise for small values of $\omega$ as well as in a narrow interval in the vicinity of $\omega=1$, a wide parametric interval of frequencies exists where the solitary waves are dynamically stable.

Finally, we have considered the dynamics of prototypical unstable solutions. To this aim, we have included the spectral plane of the corresponding solution and a link to the movie with the evolution.

\begin{figure}
\begin{center}
\begin{tabular}{cccc}
\includegraphics[width=4.5cm]{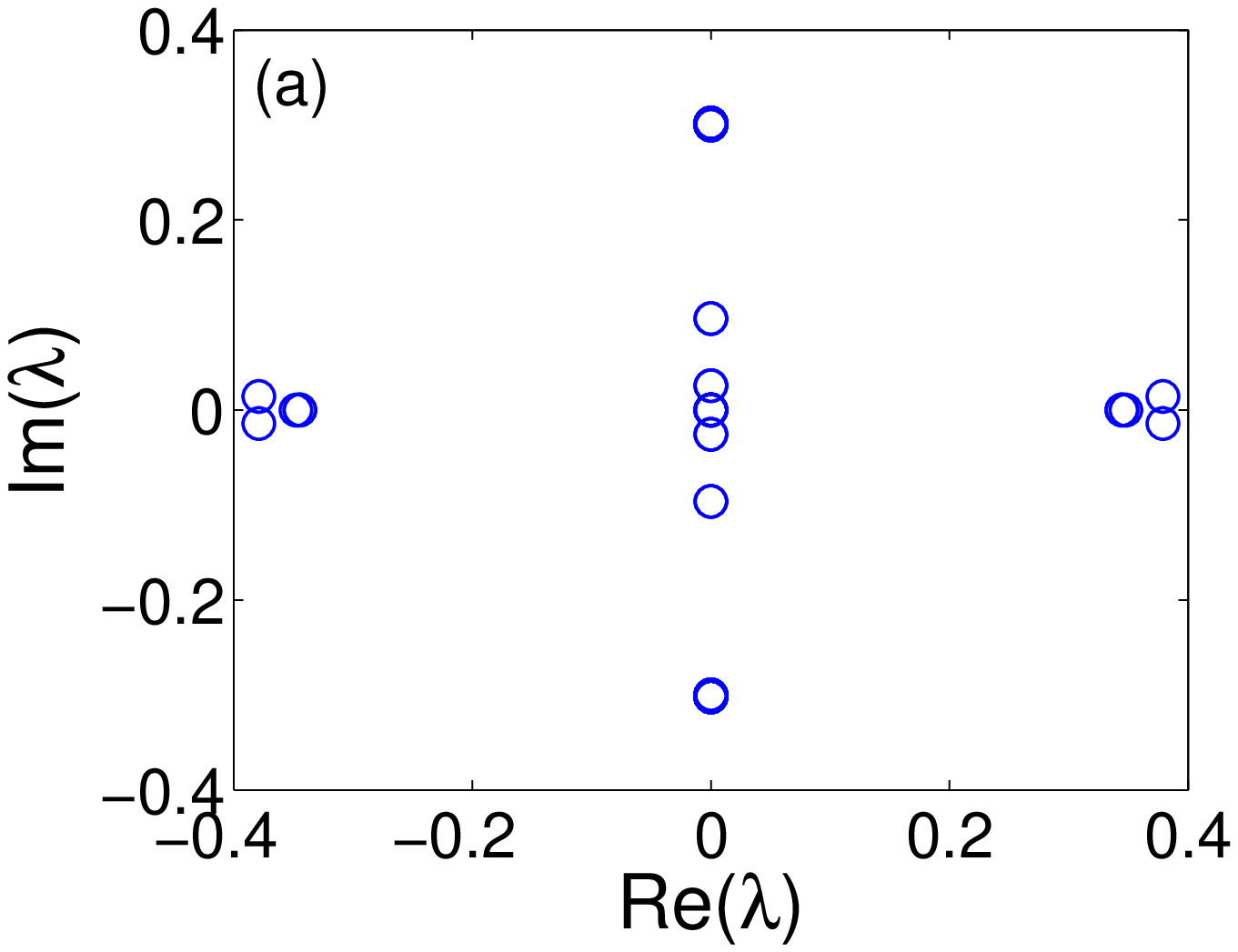} &
\includegraphics[width=4.5cm]{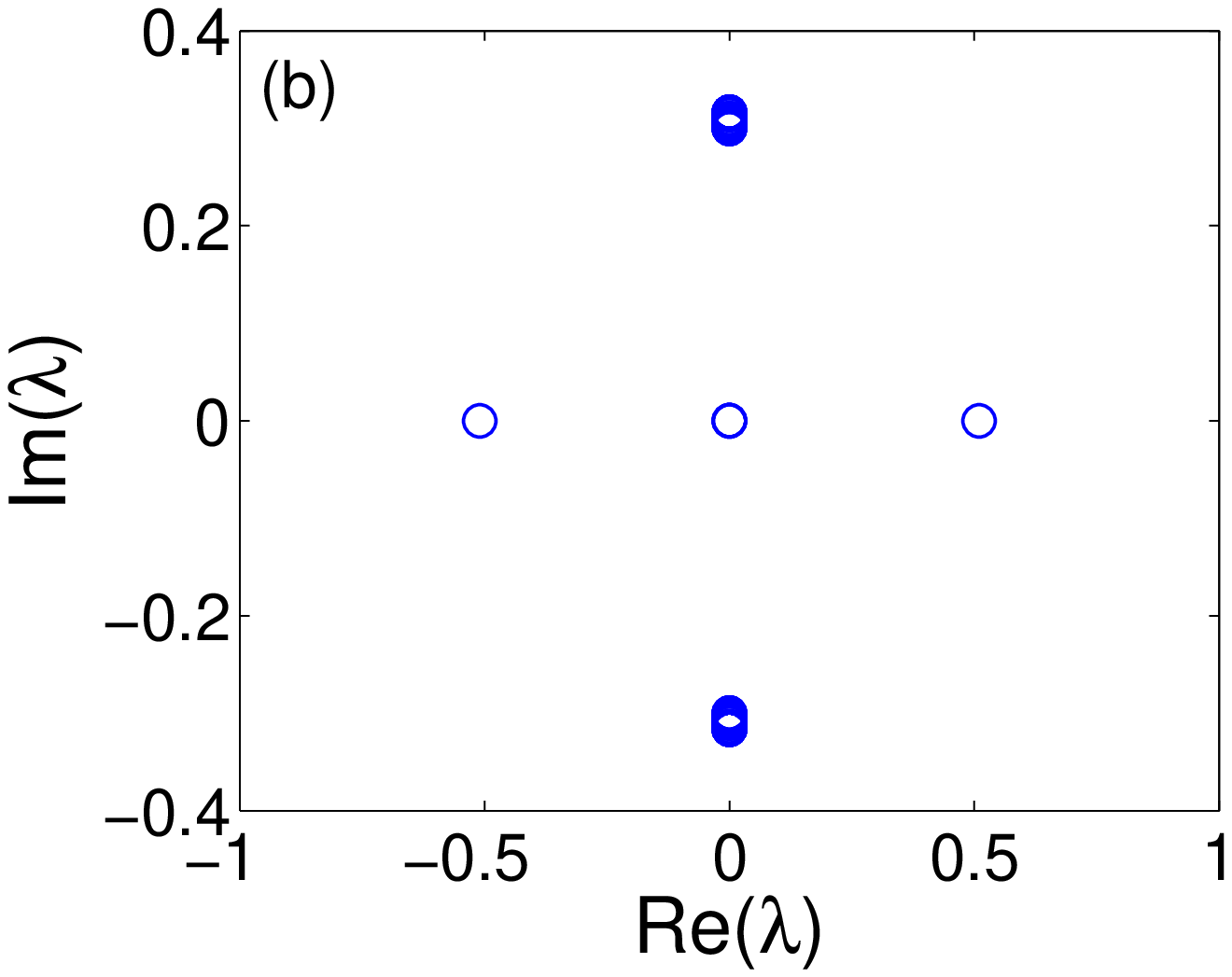} &
\includegraphics[width=4.5cm]{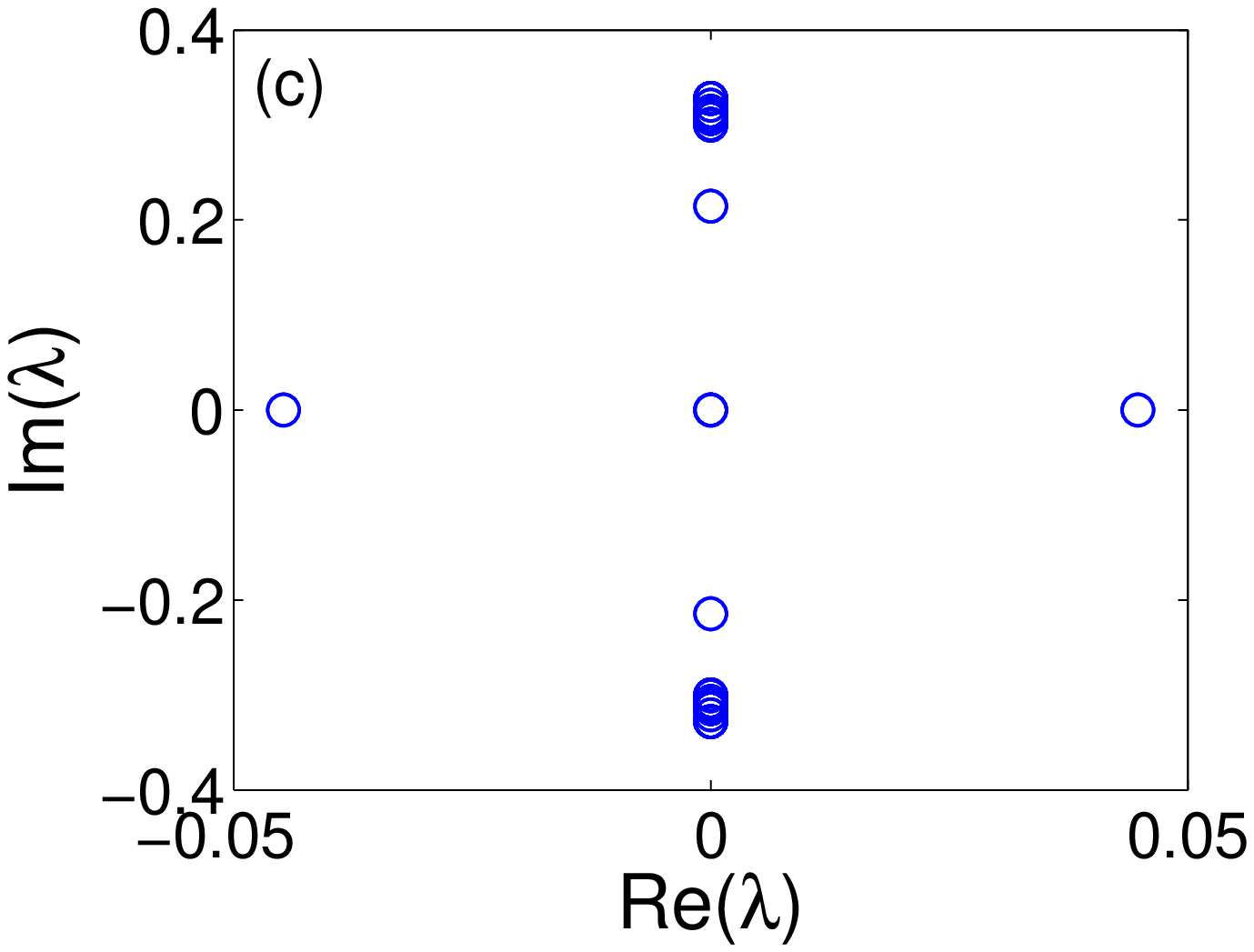} &
\includegraphics[width=4.5cm]{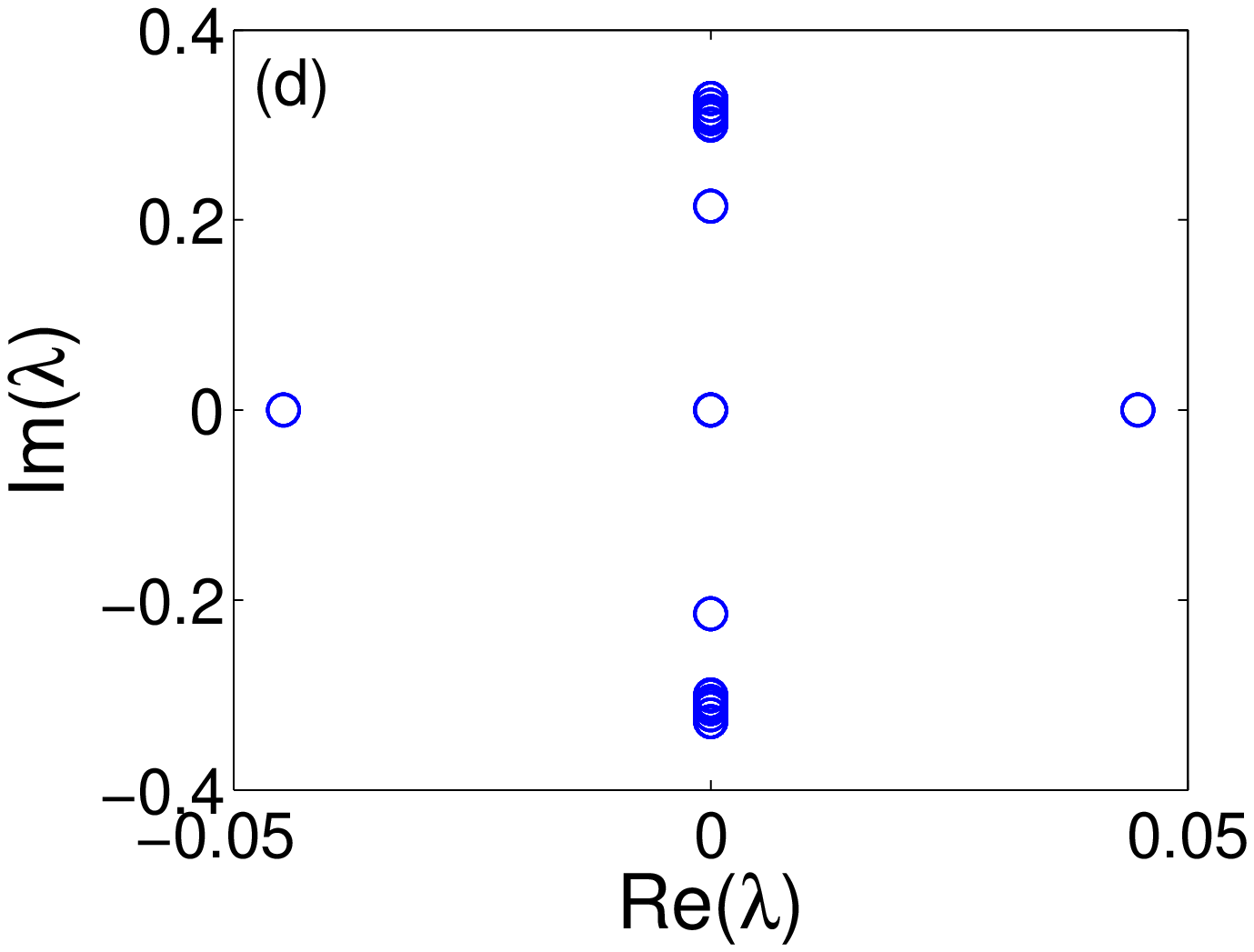} \\
\includegraphics[width=4.5cm]{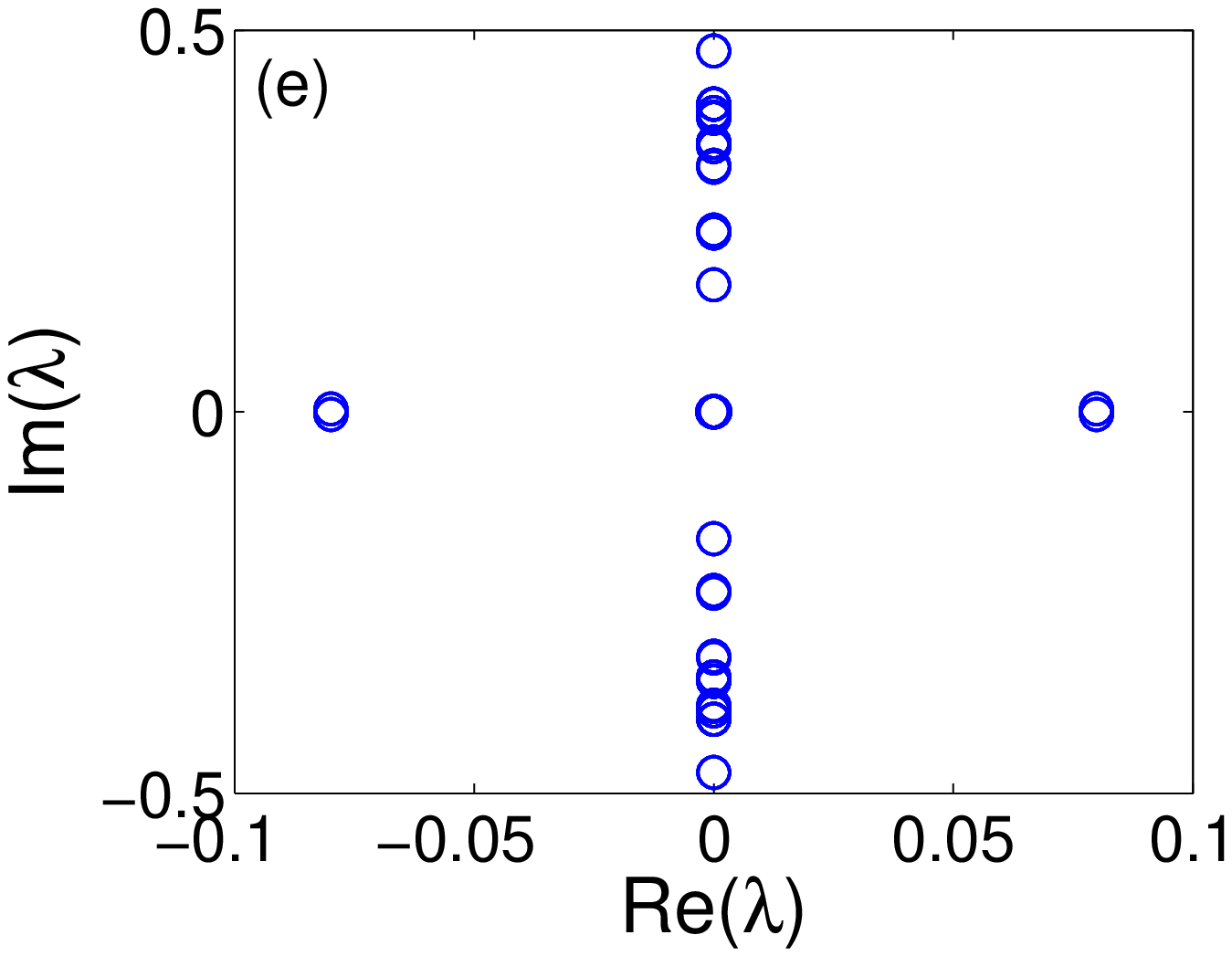} &
\includegraphics[width=4.5cm]{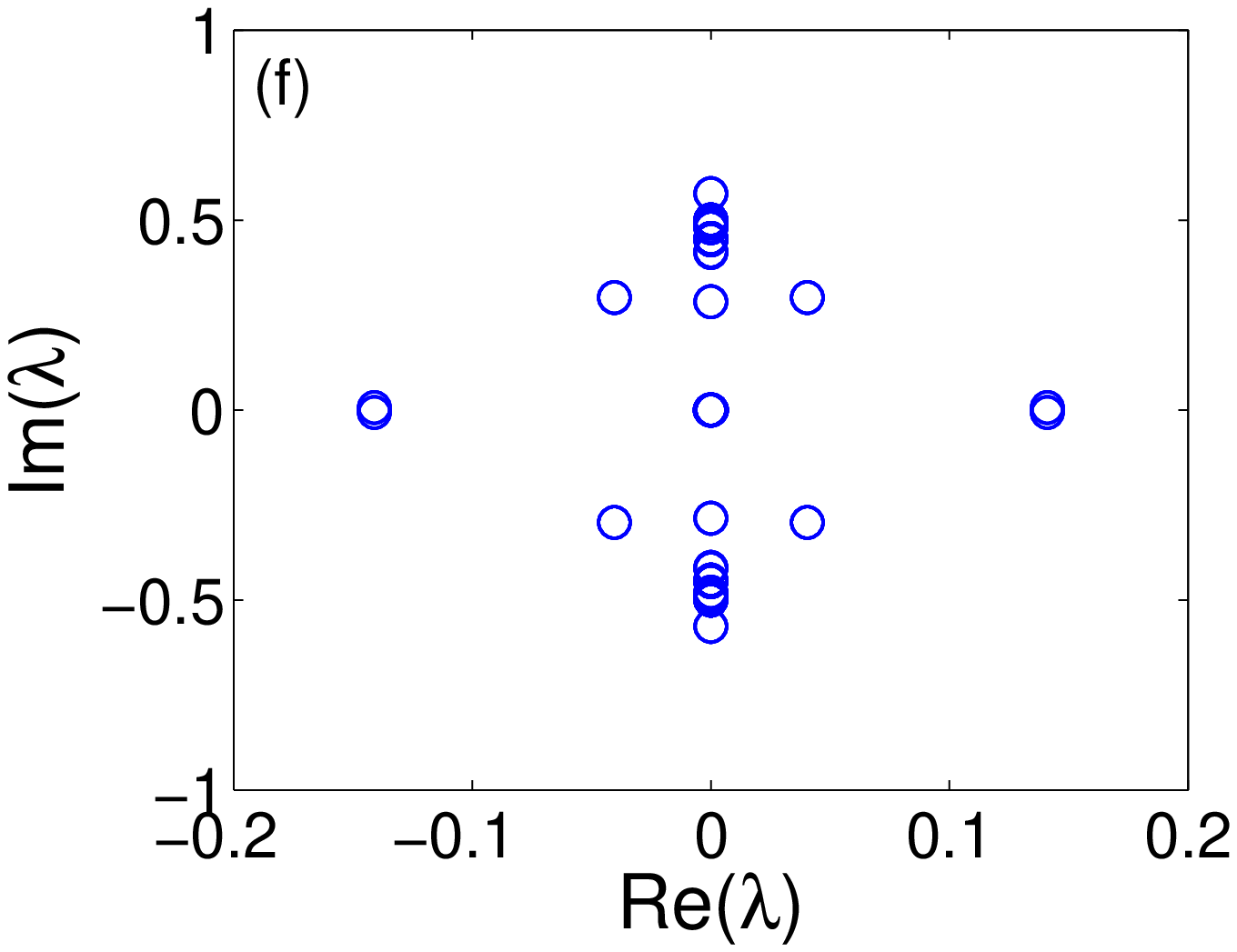} &
\includegraphics[width=4.5cm]{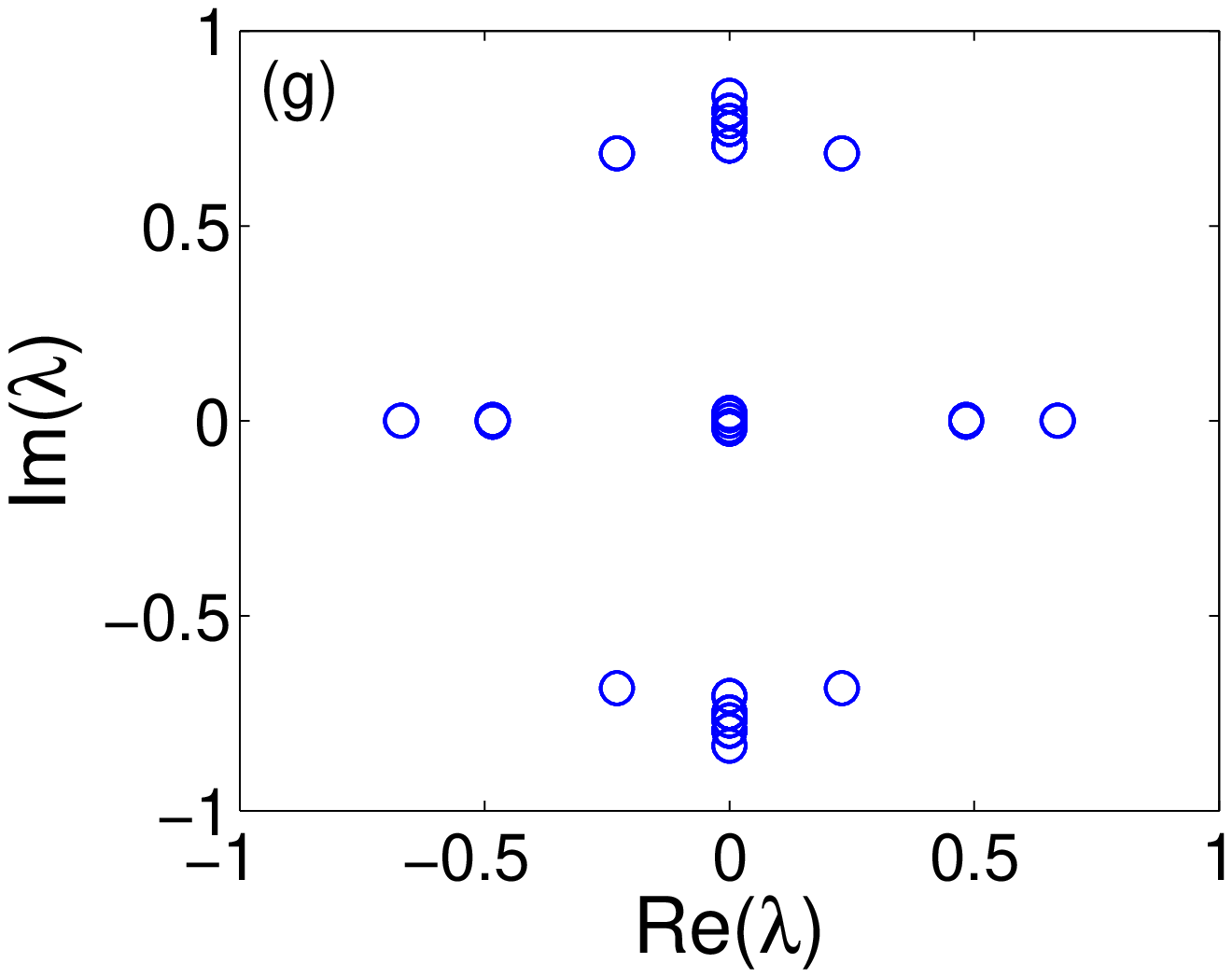} &
\includegraphics[width=4.5cm]{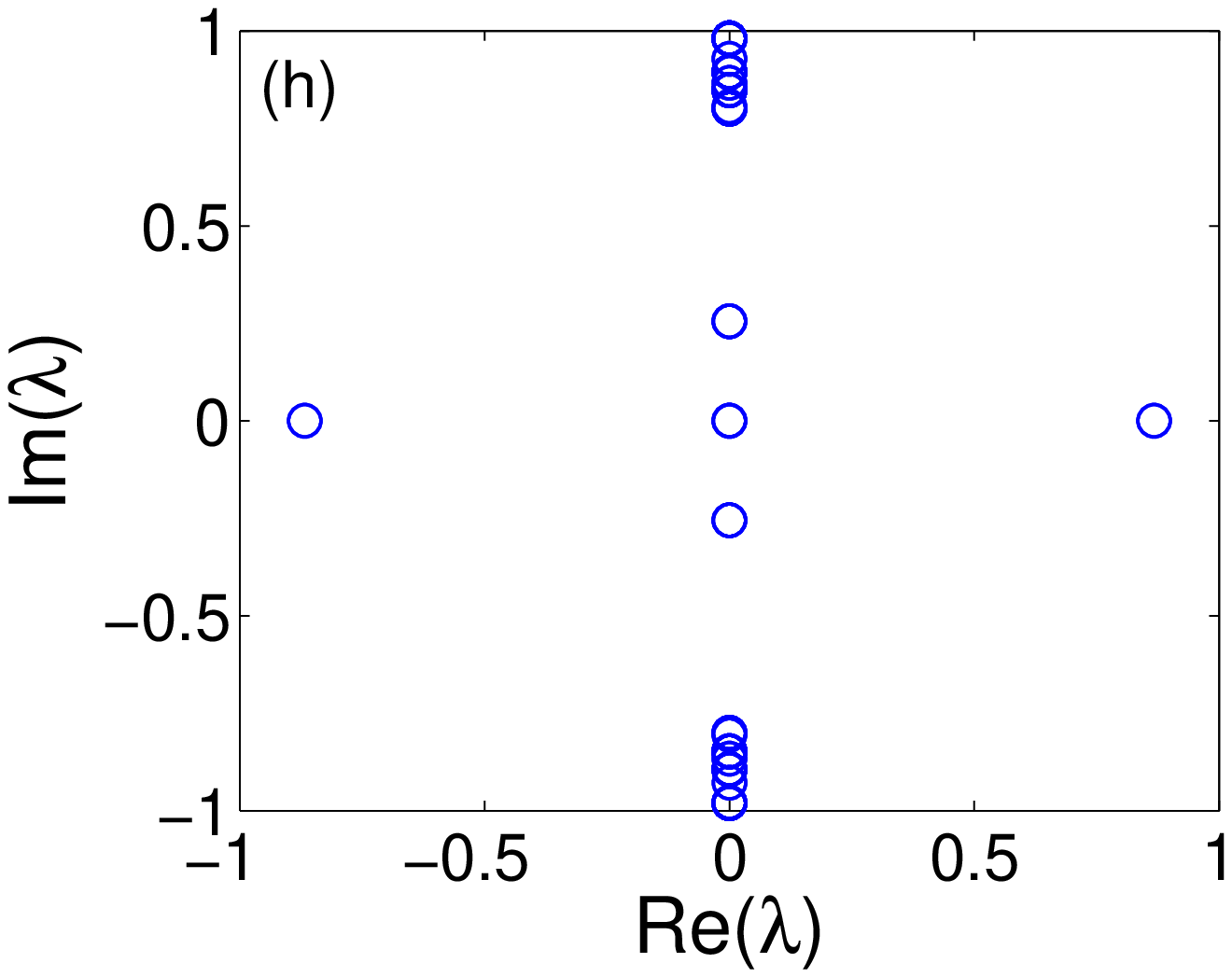} \\
\end{tabular}
\end{center}
\caption{Spectral planes of unstable solitons for which dynamics is analyzed. (a-d) correspond to fixed $\omega=0.7$ and different $C$ whereas (e-h) hold for fixed $C=5$ and variable $\omega$. (a) 9-site soliton with $C=0.5$; (b) 1-site soliton with $C=1.45$; (c) 1-site soliton with $C=1.9$; (d) 2-site soliton with $C=1.9$; (e) 9-site soliton with $\omega=0.7$; (f) 9-site soliton with $\omega=0.6$; (g) 9-site soliton with $\omega=0.3$; (h) 1-site soliton with $\omega=0.2$.}
\label{fig:dyn}
\end{figure}

We start with the typical evolution for the 9-site soliton at small coupling, where it possesses several coexisting instabilities (see Fig.~\ref{fig:dyn}a). The evolution for $C=0.5$ and $\omega=0.7$, shown in \cite{movie2}, leads to the destruction of the structure. In the case of 1-site solitons, we consider two cases corresponding to each exponential instability (Figs.~\ref{fig:dyn}b-c); the first one leads to the expansion i.e. dispersion of the solitary wave (\cite{movie3}, for $\omega=0.7$ and $C=1.45$), and the second one to slow soliton motion (\cite{movie4}, for $\omega=0.7$ and $C=1.9$). For the 2-site soliton, where the spectral dependence on $C$ is qualitatively similar to the 1-site soliton, the dynamics for the first instability is similar to the 1-site case (i.e. the soliton expands with time); however, for the second instability (see spectrum in Fig.~\ref{fig:dyn}d) we observe that the 2-site soliton breaks up (\cite{movie5}, for $\omega=0.7$ and $C=1.9$), repartitioning its mass into predominantly single site structures.

Another interesting regime for the dynamical observation of the solutions' instability regards the setting close to the continuum limit. To this end, we have fixed $C=5$ and observed the 9-site soliton dynamics for three cases (see Figs.~\ref{fig:dyn}e-g), which are traced in \cite{movie6} ($\omega=0.7$), \cite{movie7} ($\omega=0.6$) and \cite{movie8} ($\omega=0.3$). In the first case, there is only an oscillatory instability that leads to the transformation of the soliton in a pair of precessing (and periodically recombining) solitary structures; in the second case, there is an additional oscillatory instability whose consequence is the eventual destruction of the solitary wave after it is initially converted into a 1-site soliton pair. Finally, in the third case, apart from the two oscillatory instabilities identified in this case, there is a dominant exponential instability which leads to the soliton's expansion and subsequent pulsation. An example of the instability of the 1-site soliton as we approach the continuum limit (for large $C=5$ and $\omega=0.2$) is shown in~\cite{movie9}. Here the soliton ends up breathing as a result of its exponential instability.

\section{Conclusions \& Future Challenges}

In the present work, we have given a systematic account of
some of the prototypical solitary states that a two-dimensional
model of a Dirac type can bear as stationary solutions.
Our work was motivated by models of binary waveguide
lattices that have recently appeared in the literature~\cite{Tran,akyl};
thus, we have considered nonlinearities that are onsite
in each component. We have explored the model from two complementary
perspectives. We have identified the continuum limit solution and
extended it all the way to the anti-continuum limit where somewhat
surprisingly we have found it to correspond to a 9-site solution.
On the other hand, we have constructed some of the simplest
solutions of the anti-continuum limit, such as the 1- and 2-site
ones and extended them over all couplings towards the continuum limit.
In addition to the existence problem, we have provided a road map
towards the corresponding stability properties. The 1- and 2-site
solutions with their staggered structure appear to be rather
robust and, in the exception of some finite intervals of instability,
appear to feature stable dynamics. On the other hand, the 9-site
solution contains considerably more directions of potential
instability, yet most of these disappear in the large coupling regime.
The unstable dynamics of the different waveforms were also considered showing
examples of breathing, mobility, and fission as potential manifestations
of the instability, depending on the solution of interest and its
specific (frequency and coupling) parameters.

Finally, the current study suggests a number of future directions
of interest. In the context of the discrete version of the
nonlinear Schr{\"o}dinger equation, a systematic
perturbative analysis was developed from the anti-continuum limit
that enabled a characterization of the stability features
in the vicinity of this limit and the development of an understanding
of the conditions under which structures near this limit might
be stable~\cite{dnlsbook}. A similar theory seems to be within
reach in the case of the Dirac model (see also~\cite{yannan} for
a recent analysis in 1D binary waveguide arrays), but has
not been developed as of yet. On the other hand, and although
it is less relevant to the optical problem per se, extending
Dirac-like models and associated consideration to three-dimensional
settings would be a particularly challenging theme of work. Here,
once again the continuum limit preliminary conclusions of~\cite{PRL}
suggest possible existence of stable solutions, which may in principle
be possible to continue between the continuum and anti-continuum limit
and be spectrally stable in wide parametric intervals between these
two limits. Studies along these directions are currently in progress
and will be reported in future publications.

\vspace{5mm}

{\it Acknowledgements}. P.G.K. gratefully acknowledges the support of NSF-PHY-1602994, the Alexander von Humboldt Foundation, the Stavros Niarchos Foundation via the Greek Diaspora Fellowship Program,  and the ERC under FP7, Marie Curie Actions, People, International Research Staff Exchange Scheme (IRSES-605096). J.C.-M. thanks the European Regional Development Funds program (EU-FEDER) and the MEIC (project MAT2016-79866-R) for financial support. This work was supported in part by the U.S. Department of Energy.


\begin{thebibliography}{99}

\bibitem{dnc} D. N.\ Christodoulides,
F.\ Lederer, and Y.\ Silberberg,
Nature \textbf{424}, 817 (2003); A. A.\ Sukhorukov,
Y. S.\ Kivshar, H. S.\ Eisenberg, and Y.\ Silberberg,
IEEE J. Quant. Elect. \textbf{39}, 31 (2003).

\bibitem{moti} F. Lederer, G. I. Stegeman, D. N. Christodoulides, G. Assanto,
M. Segev, and Y. Silberberg, Phys. Rep. {\bf 463}, 1 (2008).

\bibitem{yaron} H. S. Eisenberg, Y. Silberberg, R. Morandotti, A. R. Boyd, and J. S. Aitchison
Phys. Rev. Lett. {\bf 81}, 3383 (1998).

\bibitem{yaron1} H. S. Eisenberg, Y. Silberberg, R. Morandotti, and J. S. Aitchison
Phys. Rev. Lett. {\bf 85}, 1863 (2000).

\bibitem{christo2} R. Iwanow, D. A. May-Arrioja, D. N. Christodoulides,
G. I. Stegeman, Y. Min, and W. Sohler,
Phys. Rev. Lett. {\bf 95}, 053902 (2005).

\bibitem{kip} C. E. R{\"u}ter, K. G. Makris, R. El-Ganainy,
D. N. Christodoulides, M. Segev, and D. Kip,
Nature Phys. {\bf 6}, 192 (2010).

\bibitem{yaron2} R. Morandotti, U. Peschel, J. S. Aitchison, H. S. Eisenberg, and Y. Silberberg
Phys. Rev. Lett. {\bf 83}, 2726  (1999);
R. Morandotti, H. S. Eisenberg, Y. Silberberg, M. Sorel, and J. S. Aitchison
Phys. Rev. Lett. {\bf 86}, 3296 (2001).

\bibitem{neshev} D. N. Neshev, T. J. Alexander,
E. A. Ostrovskaya, Yu. S. Kivshar, H. Martin, I. Makasyuk,
and Z. Chen,
Phys. Rev. Lett. {\bf 92}, 123903 (2004).

\bibitem{fleischer} J. W. Fleischer, G. Bartal,
O. Cohen, O. Manela, M. Segev, J. Hudock, and D. N. Christodoulides,
Phys. Rev. Lett. {\bf 92}, 123904 (2004).

\bibitem{dncm} J. Meier, J. Hudock, D. Christodoulides,
G. Stegeman, Y. Silberberg, R. Morandotti, and J.S. Aithcison,
Phys. Rev. Lett. {\bf 91}, 143907 (2003).

\bibitem{rudy} R. L. Horne, P. G. Kevrekidis, and N. Whitaker,
Phys. Rev. E {\bf 73}, 066601 (2006).

\bibitem{stege} R. Iwanow, R. Schiek, G. I. Stegeman, T. Pertsch, F. Lederer, Y. Min, and W. Sohler,
Phys. Rev. Lett. {\bf 93}, 113902 (2004).

\bibitem{rudy2} H. Susanto, R. L. Horne, N. Whitaker, and P. G. Kevrekidis
Phys. Rev. A {\bf 77}, 033805 (2008).

\bibitem{shandarov} E. Smirnov, C.E. R{\"u}ter, M. Stepi{\'c},
D. Kip, and V. Shandarov, Phys. Rev. E {\bf 74}, 065601 (2006).

\bibitem{hadi} E. P. Fitrakis, P. G. Kevrekidis, H. Susanto, and D. J. Frantzeskakis
Phys. Rev. E {\bf 75}, 066608 (2007).


\bibitem{ober} O. Morsch and M. Oberthaler, Rev. Mod. Phys. {\bf 78}, 179
(2006).

\bibitem{konotop1} V.A. Brazhnyi and V.V. Konotop,
Mod. Phys. Lett. B {\bf 18}, 627 (2004).

\bibitem{Tran} T.X. Tran, X.N. Nguyen, and F. Biancalana, Phys. Rev. A {\bf 91}, 023814 (2008).


\bibitem{akyl} M. Conforti, C. De Angelis, and T.R. Akylas,
Phys. Rev. A {\bf 83}, 043822 (2011).

\bibitem{jesuskip} A. Kanshu, C.E. R{\"u}ter, D. Kip, J. Cuevas,
and P.G. Kevrekidis,
Eur. Phys. J. D {\bf 66}, 182 (2012).

\bibitem{Aceves} M. Conforti, C. De Angelis, T.R. Akylas, and
  A.B. Aceves,
  Phys. Rev. A {\bf 85}, 063836 (2012).

\bibitem{yannan}
  Y. Shen, P.G. Kevrekidis, G. Srinivasan, and A. Aceves,
  J. Phys. A: Math. Theor. {\bf 49}, 295205 (2016).


\bibitem{Carr1} L. Haddad, K. O'Hara, and L.D. Carr, Phys. Rev. A {\bf 91}, 043609 (2015).

\bibitem{Carr2} L. Haddad and L.D. Carr. New J. Phys. {\bf 17}, 113011 (2015).

\bibitem{diracmat} T. O. Wehling, A. M. Black-Schaffer, and A. V. Balatsky, Adv. Phys. {\bf 63}, 1 (2014).

\bibitem{tmdc} K. F. Mak, C. Lee, J.  Hone, J. Shan, and T. F. Heinz, Phys. Rev. Lett. {\bf 105}, 136805 (2010).

\bibitem{peleg} O. Peleg,  G. Bartal, B. Freedman, O. Manela, M. Segev,
and D.N. Christodoulides,
Phys. Rev. Lett. {\bf 98}, 103901 (2007).


\bibitem{ablowitz3} M. J. Ablowitz, S.D. Nixon, and Y. Zhu,
{Phys. Rev. A} {\bf 79},  053830 (2009).

\bibitem{ablowitz4} M. J. Ablowitz and Y. Zhu,
{Phys. Rev. A} {\bf 82},  013840 (2010).

\bibitem{PRL} J. Cuevas-Maraver, P.G. Kevrekidis, A. Saxena, A. Comech and R. Lan, Phys. Rev. Lett. {\bf 116}, 214101 (2016).

\bibitem{JPA} J. Cuevas-Maraver, P.G. Kevrekidis, and A. Saxena, J. Phys. A: Math. Theor. {\bf 48}, 055204 (2015).


\bibitem{macaub} R.S. MacKay and S. Aubry,
\newblock Nonlinearity {\bf 7}, 1623 (1994).

\bibitem{movie1} \url{https://www.dropbox.com/s/u5pur2adm81dfjz/movie1.avi?dl=0} (See also movie1.avi in Supplementary Material)

\bibitem{movie2} \url{https://www.dropbox.com/s/1e5yqhw9c9g30ef/movie2.avi?dl=0} (See also movie2.avi in Supplementary Material)

\bibitem{movie3} \url{https://www.dropbox.com/s/u3j6p20opx9hbpm/movie3.avi?dl=0} (See also movie3.avi in Supplementary Material)

\bibitem{movie4} \url{https://www.dropbox.com/s/1zmh73vwh4w9liz/movie4.avi?dl=0} (See also movie4.avi in Supplementary Material)

\bibitem{movie5} \url{https://www.dropbox.com/s/zl2nqe5c5qxg7lj/movie5.avi?dl=0} (See also movie5.avi in Supplementary Material)

\bibitem{movie6} \url{https://www.dropbox.com/s/66mtzvmyx1xynms/movie6.avi?dl=0} (See also movie6.avi in Supplementary Material)

\bibitem{movie7} \url{https://www.dropbox.com/s/wg1i0vw7n4vnmf6/movie7.avi?dl=0} (See also movie7.avi in Supplementary Material)

\bibitem{movie8} \url{https://www.dropbox.com/s/od5xp5nbz2mtkcm/movie8.avi?dl=0} (See also movie8.avi in Supplementary Material)

\bibitem{movie9} \url{https://www.dropbox.com/s/glmyr8044mc0gvi/movie9.avi?dl=0} (See also movie9.avi in Supplementary Material)

\bibitem{dnlsbook} P.G. Kevrekidis,
{\it The Discrete Nonlinear Schrödinger Equation},
Springer-Verlag (Heidelberg, 2009).



\end{thebibliography}
\end{document}